\documentclass[12pt]{article}
\usepackage{graphicx}
\usepackage{latexsym}
\advance\oddsidemargin by -\textwidth
\advance\oddsidemargin by -1in
\evensidemargin=\oddsidemargin
\def\fsl#1{\setbox0=\hbox{$#1$}           % set a box for #1 
   \dimen0=\wd0                                 % and get its size
   \setbox1=\hbox{/} \dimen1=\wd1               % get size of /
   \ifdim\dimen0>\dimen1                        % #1 is bigger
      \rlap{\hbox to \dimen0{\hfil/\hfil}}      % so center / in box
      #1                                        % and print #1
   \else                                        % / is bigger
      \rlap{\hbox to \dimen1{\hfil$#1$\hfil}}   % so center #1
      /                                         % and print /
   \fi}                                         %
\makeatletter
\def\@maketitle{\newpage
 \vspace*{-2cm}
 \null
 {\normalsize \tt \begin{flushright} 
  \begin{tabular}[t]{l} DPNU-05-15\\ TU-751\\ YITP-SB-05-28 %\@date 
  \end{tabular}
 \end{flushright}}
 \begin{center}
 \vskip 2em
 {\LARGE \@title \par} \vskip 2.5em {\large \lineskip .5em
 \begin{tabular}[t]{c}\@author 
 \end{tabular}\par} 
 \end{center}
} 

\newcommand{\vereq}[2]{\lower3pt\vbox{\baselineskip1.5pt \lineskip1.5pt
\ialign{$\m@th#1\hfill##\hfil$\crcr#2\crcr\sim\crcr}}}
\newcommand{\lesssim}{\mathrel{\mathpalette\vereq<}}

\newcommand\eqsecnum{
\@newctr{equation}[section]
\renewcommand\theequation{\arabic{section}.\arabic{equation}}%
}

\newbox\tempboxa
\newdimen\captionboxsubcount 
\def\capsize#1{\captionboxsubcount=#1pt}
\newdimen\captionboxsub
\captionboxsub=\hsize \advance\captionboxsub by -\captionboxsubcount
\advance\captionboxsub by -\captionboxsubcount
\long\def\@makecaption#1#2{
 \setbox\@tempboxa\hbox{\footnotesize #1: #2}
 \ifdim \wd\@tempboxa >\captionboxsub 
\rightskip=\captionboxsubcount \leftskip=\captionboxsubcount 
  \footnotesize #1: #2 
\else \hbox to\hsize{\hfil\box\@tempboxa\hfil} 
 \fi}
\makeatother
\eqsecnum
\capsize{30}
\title{{\Large\bf 
 $\pi^+ - \pi^0$ mass difference and
$S$ parameter 
in the
   large $N_f$ QCD\thanks{A preliminary report~\cite{Harada:2005zz} was given at the 2004 International
Workshop on Dynamical Symmetry Breaking (DSB04), December 21-22, 2004, 
Nagoya Univeristy, Nagoya.}
\vspace{3mm}}}
\author{
{\large
    Masayasu {\sc Harada$^{\,a,\,}$}%\thanks{
     \ ,\ 
    Masafumi {\sc Kurachi$^{\,b,\,}$}%\thanks{
    \ \  and \ \ 
    Koichi {\sc Yamawaki$^{\,a,\,}$}%\thanks{
  }
 \\[5mm]
  {\it $^{a} $Department of Physics, Nagoya University}\\
  {\it Nagoya 464-8602, Japan}\\[4mm]
  {\it $^{b} $Department of Physics, Tohoku University}\\
  {\it Sendai 980-8578, Japan}
  \\
  and\\
{\it C.N. Yang Institute for Theoretical Physics, State University of New York,}
\\
{\it Stony Brook, NY 11794, U.S.A.}
 \thanks{Present address}
 \\[1cm]
}

\begin{document}

\maketitle

%%%%%%%%%%%%%%%%%%%%%%%%%%%%%%%%%%%%%%%%%%%%%%%%%%%%%%%%%%%%%%%%%%%%
%%%%%%%%%%%%%%%%%%%%%%%%%%%%%%%%%%%%%%%%%%%%%%%%%%%%%%%%%%%%%%%%%%%%
\vspace{-1.1cm}
\begin{abstract}
In the framework of the Schwinger-Dyson equation and the Bethe-Salpeter equation in the 
improved ladder approximation, 
we calculate 
the $S$ parameter and an analogue of the $\pi^+ - \pi^0$ mass difference $\Delta m_\pi^2
 \equiv  m_{\pi^+}^2 -m_{\pi^0}^2$ 
as well as the NG boson decay constant $f_\pi$ 
on the same footing in the large $N_f$ QCD, 
through the difference between the vector current correlator
$\Pi_{VV}$ and the axial-vector current correlator $\Pi_{AA}$.
Approaching the chiral phase transition point 
$\alpha_*\rightarrow \alpha_{\rm cr} (=\pi/4)$ from the broken phase,
where $\alpha_*$ is the the gauge coupling on the infrared fixed point,
$\Delta m_\pi^2$ as well as $f_\pi^2$ goes to zero with the essential-singularity 
scaling (Miransky scaling), while the ratio indicates a blowing up enhancement 
reflecting  the characteristic behavior of the large $N_f$ QCD as a walking theory
which is expected to scale as 
%consistent 
%with the inverse square root scaling,  
$\Delta m_\pi^2/f_\pi^2 \sim 
\left(\alpha_*/\alpha_{\rm cr}-1 \right)^{-1/2}.
$
%This enhancement is  a new effect 
%which we find is a direct consequence of  
%but has been overlooked in the 
%discussions of the walking/conformal (scale-invariant) technicolor.
On the other hand,
the $S$ parameter 
takes values somewhat smaller 
than that of the real-life QCD and indicates slightly decreasing tendency %,
%with $\left(\Delta m_\pi^2/f_\pi^2\right) \cdot S$ being kept constant
as we approach the phase transition point.
%, which is consistent with
%a  square root scaling,  $S \sim 
%\left(\alpha_*/\alpha_{\rm cr}-1\right)^{1/2}$, implying that 
%$S$ parameter does vanish near the critical point.
\end{abstract}

\newpage
%%%%%%%%%%%%%%%%%%%%%%%%%%%%%%%%%%%%%%%%%%%%%%%%%%%%%%%%%%%%%%%%%%%%
%%%%%%%%%%%%%%%%%%%%%%%%%%%%%%%%%%%%%%%%%%%%%%%%%%%%%%%%%%%%%%%%%%%%
\section{Introduction}
\label{sec:Introduction}

The electroweak symmetry breaking in the standard model (SM) 
is well described by introducing the Higgs boson.
However, the Higgs sector in the SM cannot {\it explain} the origin 
of the electroweak symmetry breaking, which means the SM cannot explain
the origin of mass. 
Moreover, the mass of the Higgs particle has to be fine tuned
since it receives quantum correction which is proportional to the square
of the cutoff scale $\Lambda$.
These facts naturally lead us to the idea that the Higgs sector in the 
SM is nothing but an low energy effective theory of some underlying
theory. To consider the existence of a new strong dynamics is one
example of this idea. 

The ``large $N_f$ QCD'', a jargon of the $SU(3)$ gauge theory with
a large number of massless fermions ($N_f \lesssim\frac{11}{2}N_c$)
~\cite{Banks:nn,Appelquist:1996dq, 
Miransky:1996pd,lattice,OZ,VS,Harada:1999zj,Harada:2000kb,Gies:2005as},
is 
one of the candidates for 
such a strong dynamics like the technicolor \cite{TC, techni}
which breaks the electroweak symmetry. Actually, it has interesting 
features:

\begin{itemize}

\item A dynamical model for the walking/conformal (scale-invariant) technicolor\\
It was observed~\cite{Banks:nn} that QCD has an infrared (IR) fixed point
$\alpha =\alpha_*$ for a large number of massless
fermions ($N_f \lesssim\frac{11}{2}N_c$) at the two-loop beta function.
It was regarded~\cite{Appelquist:1996dq,Miransky:1996pd} as a good 
example of 
the walking technicolor~\cite{Yamawaki:1985zg}, because the running 
gauge coupling 
near the infrared fixed point is almost constant, $\alpha(Q^2)  
\simeq \alpha_*$, namely ``walking'' for a wide range of the momenta $Q$. 

Moreover, it was found~\cite{Appelquist:1996dq}
through the improved ladder Schwinger-Dyson (SD) equation that chiral
symmetry restoration takes place for certain $N_f$ such that the IR fixed point becomes
less than the critical coupling which is determined by the SD equation, 
namely $N_f^{\rm crit} <
N_f < \frac{11}{2} N_c$, where $N_f^{\rm crit}\simeq 4 N_c$($=12$ for
$N_c=3$). 
In Ref.~\cite{Miransky:1996pd} this chiral phase transition  at
$N_f^{\rm crit}$ was further identified with the ``conformal phase
transition'' which was characterized by the essential singularity
scaling (Miransky scaling).  
 Moreover, 
the chiral restoration in large $N_f$ QCD 
was also observed in lattice simulations~\cite{lattice} and by other various
methods such as the dispersion
relation~\cite{OZ}, instanton calculus~\cite{VS}, effective field
theoretical approach~\cite{Harada:1999zj}, 
renormalization group 
flow equations~\cite{Gies:2005as}, perturbative calculus~\cite{Ndili:2005ni}
, etc., although the value of 
$N_f^{\rm crit} $
is raging from $N_f^{\rm crit}\sim 5$ to $N_f^{\rm crit} 
\sim 12$ ($N_c=3$) depending on the approaches.

\item  A model for the electroweak baryogenesis\\
The electroweak baryogenesis~\cite{Kuzmin:1985mm} (for a review see, e.g., Ref.\cite{Cohen:1993nk}) is an attractive scenario 
but has serious difficulties: 
1) lack of mechanism to produce the first order phase transition consistently
with the present experimental lower bound of the Higgs boson mass and 
2) smallness of the baryon asymmetry if the CP violation is solely due to 
the KM phase in the Standard Model.
The large $N_f$ QCD when applied to the technicolor
has new ingredients to solve these problems:
It is widely believed~\cite{Pisarski:1983ms} that QCD phase transition is the 
first order when the number
of massless flavors exceeds 3 and, we expect, so is
the chiral phase transition in the large $N_f$ QCD.
Moreover, it was argued~\cite{Martin:2004ec} that 
the technicolor may have extra sources of the CP 
violation other than the KM phase and supply enough amount of CP violation 
needed for the baryogenesis.
\end{itemize}

Besides application to the technicolor, much attention has been paid to the  property of the phase transition
not just the existence of the chiral phase transition in the large $N_f$ QCD.  
Especially, it is interesting to ask what are the light degrees of
freedom near the phase transition point in the large $N_f$ QCD: 
For example, in the manifestation of the chiral symmetry restoration  
\'{a} la 
linear sigma model (Ginzburg-Landau type effective theory), 
the scalar bound state becomes a chiral
partner of the pseudoscalar bound state and becomes massless at the
phase transition point when approached both from the broken and the symmetric phases. However, it was emphasized from the viewpoint of the conformal phase
transition~\cite{Miransky:1996pd} that the Ginzburg-Landau
type effective theory breaks down in the approach of the
improved ladder SD equation, signaled by the absence of massless scalar bound state 
in the symmetric phase. Quite recently, 
such a peculiarity of the phase transition
was also observed based on the renormalization-group 
analysis~\cite{Gies:2005as}.
Besides, from the viewpoint of the conformal phase
transition
, it is natural to suppose that all
the 
existing bound states become massless near the phase transition point
when approached from the broken phase (see
Ref.~\cite{Chivukula:1996kg}). 
On the other hand, in the vector 
manifestation~\cite{Harada:2000kb} obtained by the effective field
theoretical approach based on the hidden local symmetry
theory~\cite{BKUYY}, it is the vector bound state
which 
becomes massless as a chiral partner of the pseudoscalar bound state.
It remains unclear at this moment whether or not the vector manifestation contradicts the
conformal phase transition.

In Ref.~\cite{Harada:2003dc}, in the framework of the SD and the homogeneous
Bethe-Salpeter (HBS) equations (see 
\cite{Kugo:1991da} for a review),
we studied 
which types of the light bound 
states actually exist near the phase transition point, and
investigated 
the critical behavior of their masses directly from QCD.
We found that approaching the chiral phase transition point from
the broken phase, the scalar, vector, and axial-vector meson masses
vanish to zero with the same scaling behavior, all degenerate with the
massless pseudoscalar meson.

In this paper, we further investigate the properties of the chiral
phase transition in the large $N_f$ QCD 
through the critical behavior
of the $S$ parameter 
and the ``$\pi^+ - \pi^0$ mass
difference $\Delta m_\pi^2 \equiv m_{\pi^+}^2 -m_{\pi^0}^2$'',
by extending our previous work~\cite{Harada:2004qn} on the S parameter
and $\Delta m_\pi^2 $ done for the real-life QCD
with $N_c=N_f=3$. They are interesting quantities when we apply the dynamical symmetry breaking scenario based on the large $N_f$ QCD for the  
 model buildings beyond the SM like the walking technicolor, UV completions (underlying theories)
 for the
 little Higgs models \cite{Arkani-Hamed:2001nc} and the Higgsless models \cite{Csaki:2003dt}, etc.

Actually, there 
are 
strong constraints on the electroweak 
$S$ parameter~\cite{PeskinTakeuchi}
from electroweak precision tests, it is quite
important to estimate the $S$ parameter theoretically 
when we consider the technicolor-like scenario. 
Estimation of masses of pseudo NG bosons is also important to check
that 
the model is consistent with the fact that such a pseudo NG bosons
have not been discovered so far.
Then, it is desirable to estimate the $S$ parameter and 
$\Delta m_\pi^2$ reliably.

Several attempts to estimate the parameter 
${\hat S}$, which is the contribution to the $S$ parameter from one
weak fermionic doublet, for the 
 walking technicolor theories 
 and the large $N_f$ QCD
 have been made
so far by using several methods 
\cite{Sundrum:1991rf,Appelquist:1998xf}.
However, it is difficult to estimate the accuracy of these calculation
(see, for example, Ref.~\cite{TakNagoya}).
What makes it difficult to estimate the $S$ parameter is its strong
dependence on the nonperturbative dynamics of the low momentum region. 
Then it is quite important to calculate the $S$ parameter by using the
nonperturbative method 
in a way to directly
deal the fermions and the gauge bosons as the fundamental degrees of
freedom.  

As such, in the previous work~\cite{Harada:2004qn}, we calculated
 the 
 $S$ parameter and $\Delta m_\pi^2$ in the real-life ($N_f = 3$) QCD
through the difference between
the vector 
current correlator $\Pi_{VV}$ and the axial-vector current correlator
$\Pi_{AA}$.
$\Pi_{VV} - \Pi_{AA}$ were calculated in the framework of the SD
equation and the inhomogeneous Bethe-Salpeter (IBS) equation in the
improved ladder approximation.
It was stressed that these physical quantities are calculated by using
the correlators {\it in the space-like region}, so that no analytic
continuation is needed.
The results in Ref.~\cite{Harada:2004qn} shows that both the $S$
parameter and $\Delta m_\pi^2$ can be fit to the experimental values  
at the same time. 
By fitting to the calculated data using the pole
saturated form of $\Pi_{VV} - \Pi_{AA}$, we also derived masses and
decay constants of $\rho$ meson and $A_1$ meson, which we found are
consistent with the experiments.
This is quite encouraging to extend our method to the large $N_f$ QCD.

In extending the previous method to the large $N_f$ QCD,
let us consider the situation that 
$SU(N_f)_L \otimes SU(N_f)_R$ 
chiral symmetry is explicitly broken by the $U(1)$ gauge symmetry.
We take the generator of this $U(1)$ gauge symmetry as
\begin{equation}
 Q = \left(
      \begin{array}{ccc}
       2/3 &      &        \\
       & -1/3 &        \\
       &      & \ddots
      \end{array}
     \right) .
\label{def:Q}
\end{equation}
Then, we calculate the mass of the pseudo Nambu-Goldstone (NG) boson
which is associated with the following generator:\cite{Harada:2003xa}
\begin{equation}
 T_{pNGB} = \left(
      \begin{array}{cccc}
       0 & 1 & 0 & \cdots  \\
       0 & 0 & 0 &         \\
       0 & 0 & 0 &         \\
       \vdots &&& \ddots
      \end{array}
     \right) .
\label{eq:pNGB}
\end{equation}
In the case of the real-life 
QCD, the above pseudo NG boson
corresponds to the $\pi^+$ meson, and its (squared) mass caused by the
$U(1)$ electro-magnetic  gauge interaction corresponds to the $\pi^+ - \pi^0$ mass
difference $\Delta m_\pi^2 \equiv m_{\pi^+}^2 -m_{\pi^0}^2$.
Then, throughout this paper, we call the squared mass of the above
pseudo 
NG boson ``$\Delta m_\pi^2$'' even in the case of $N_f \neq 3$.

We then calculate  ${\hat S}$ and $\Delta m_\pi^2$ as well as $f_\pi$ 
at the same time.
We find that as we approach the critical point 
both $\Delta m_\pi^2$ and $f_\pi^2$  vanish in 
the essential singularity scaling (Miransky-type scaling), 
$\sim \exp
\left(
-2q
%\pi
/\sqrt{
\alpha_*/\alpha_{\rm cr} -1}
\right)
$, ($q = {\rm const.}$)
with the ratio $\Delta m_\pi^2/f_\pi^2$ being {\it much enhanced} compared with that of the real-life QCD and having a blowing up tendency 
%consistent with 
reflecting the inverse square root scaling, 
$\Delta m_\pi^2/f_\pi^2 \sim 
\left(\alpha_*/\alpha_{\rm cr}-1 \right)^{-1/2}$, which  is characteristic to %due to 
the walking gauge coupling with large anomalous dimension $\gamma_m \simeq 1$ 
of the large $N_f$ QCD. In fact the integral for the  $\Delta m_\pi^2/f_\pi^2$ 
is expected to be logarithmically divergent for the walking theory, 
$\sim \ln \left(\Lambda^2/\Lambda_\chi^2\right)$, if the walking behavior $\Pi_{VV} - \Pi_{AA} \sim 1/Q^2$
is operative for $\Lambda_\chi <Q^2 <\Lambda^2$, where $\Lambda$ is the (two-loop) scale parameter of the large $N_f$ QCD like $\Lambda_{\rm QCD}$ in the real-life QCD, and $\Lambda_\chi \simeq 4 \pi f_\pi \sim \Lambda \exp
\left(-q
%\pi
/\sqrt{\alpha_*/\alpha_{\rm cr}-1}\right)$. In the actual numerical calculations we used an artificial
cutoff $\Lambda_{\rm Num}$ which is defined for the sake of numerical calculations as 
$\Pi_{VV} - \Pi_{AA}|_{Q^2=\Lambda_{\rm Num}^2}=\frac{1}{50}\left[
\Pi_{VV} - \Pi_{AA} \right]|_{Q^2=0}$ 
%As we get closer to the critical point, the artificial cutoff  $\Lambda_{\rm Num}$ becomes 
%much smaller than the real cutoff $\Lambda$ , and hence our
%numerical estimate from $\Lambda_\chi <Q^2 <\Lambda_{\rm Num}^2$ can become  much
%smaller than the contributions from $ \Lambda_{\rm Num}^2 <Q^2 <\Lambda^2$ 
(we will later
include contributions from $\Lambda_{\rm Num}^2< Q^2 <\Lambda^2$, 
based on the analytical considerations on the walking theory).
%Nevertheless 
Up to this limitation the numerical result
%As to $\Delta m_\pi^2/f_\pi^2$, 
%Although their actual values in our calculations 
%are less than $0.6$ due to our present computational ability in the %more
%closer vicinity of 
%the critical point, they are
is already enhanced up till $\Delta m_\pi^2/f_\pi^2 \simeq 0.6$,
 about four times larger than that obtained in the same method 
for the real-life QCD, $\Delta m_\pi^2/f_\pi^2 \simeq 0.123$ ~\cite{Harada:2004qn}  (experimental value:
$\Delta m_\pi^2/f_\pi^2 =0.148\pm 0.001$), even though our calculation is still away from the
critical point for the reason of the computation ability. (If we include contributions 
 from $\Lambda_{\rm Num}^2< Q^2 <\Lambda^2$,
the enhancement is more dramatic).

This implies a {\it large enhancement factor} for the pseudo  NG 
 boson mass %which has been overlooked 
in the walking/conformal (scale-invariant) technicolor~\cite{Yamawaki:1985zg}. 
\footnote{
A similar enhancement of pseudo NG boson mass
in the walking technicolor (other than the large $N_f$ QCD) 
was also observed by a numerical study in a different 
approach~\cite{Holdom:1987ed}.
}
If it is applied to the popular one-family model
(Farhi-Susskind model)~\cite{techni}, 
the ${\rm mass}^2$ of the $P^{\pm}$ (and also of the colored pseudo NG bosons)
as a scale-up of $\Delta m_\pi^2$ essentially 
irrelevant to the ETC interactions, 
%which was
%considered not to be enhanced by the walking/conformal (scale-invariant) technicolor, 
will be enhanced
also by a large anomalous dimension
$\gamma_m \simeq 1$ by a factor four (to nine): The mass of  $P^{\pm}$ 
will be  boosted to typically more than $200(-300) {\rm GeV}$ mass range instead of $100 {\rm GeV}$
the value in a simple scale up of QCD. If we consider the model which is closer to the critical point, the
 enhancement factor should be more dramatic due to the inverse square root scaling
 mentioned above.
%On the other hand, 
%the result is somewhat different from the suggestion of 
%Ref. \cite{Harada:2003xa}
%to regard the large $N_f$ QCD as a candidate for the UV completion of the little Higgs model.

On the other hand, %contrary, 
the values of ${\hat S}$ in our calculations are
somewhat smaller  ${\hat S} = 0.25-0.30$ than that obtained in the same method for the real-life QCD
($N_c=N_f=3$), $S=\hat{S} \simeq 0.33$ ~\cite{Harada:2004qn} (experimental value: $S 
= 0.32\pm 0.04$), and indicate a tendency to decrease slightly as we approach to
the critical point. Although it is still larger than the simple one-loop result
$\hat{S}=\frac{N_c}{6 \pi}=0.16 \, (N_c=3) $, 
%In fact if we take a product $\Delta m_\pi^2/f_\pi^2\cdot \hat{S}$, it is almost constant as we vary $\alpha_*$, which suggests that
%$\hat{S}$ scales inversely to $\Delta m_\pi^2/f_\pi^2$, namely {\it scales
%to vanish as square root}, $\hat{S} \sim  \left(\alpha_*/\alpha_{\rm cr}-1 \right)^{1/2} \rightarrow 0 $ as $\alpha_*\rightarrow \alpha_{\rm cr}$, so does the $S$ parameter, $S = \frac{N_f}{2} \hat{S}$ (for $N_c=3$). Then it is easy to arrange  
%$N_f$ and $N_c$ to 
it could be  suggestive
for a further decreasing towards the critical point.  It would motivate %be interesting
further study in the points closer to the critical point to see whether or not
%and is not enough for the % to suppress the $S$ parameter in the
technicolor model based on the large $N_f$ QCD 
makes $S$ consistent with the present experimental bound $S < 0.1$.
%This will be the simplest solution to the most serious
%problem of the technicolor, since a viable model building can be made by
%simply playing with $N_f$ and $N_c$ which are at our disposal in a way we can 
%easily incorporate the technicolor sector into a larger picture like a  
%simple ETC-type models.  

The calculation in this paper turns out to be the first example which
estimates the $S$ parameter and $\Delta m_\pi^2$ 
in the large $N_f$ QCD as a model of walking 
technicolor
by directly solving
its dynamics
\footnote{
 In Ref.~\cite{Harada:1994ni}, the $S$ parameter in the
 walking technicolor theory was calculated by using the BS equation.
 However, the $S$ parameter in the large $N_f$ QCD was not calculated.
}.

This paper is organized as follows.
In the next section, we briefly review the large $N_f$ QCD.
In section~\ref{sec:Spectral}, we introduce sum rules: The DMO sum
rule or the zeroth Weinberg sum rule for  
the parameter $\hat{S}$, the first Weinberg sum rule for $f_\pi^2$,
and the DGMLY sum rule or the third Weinberg
sum rule for $\Delta m_\pi^2$. We then rewrite them in terms of 
the current correlators $\Pi_{VV} - \Pi_{AA}$.
In section~\ref{sec:correlator-BS}, we show how the current
correlators are obtained from the BS amplitudes
calculated from the IBS equation in the space-like region.
In section~\ref{sec:IBS}, we introduce the IBS equation.
In section ~\ref{sec:results}, we show the results of numerical
calculation for critical behaviors of $f_\pi$,  
$\Delta m_\pi^2$ and  ${\hat S}$. 
Section~\ref{sec:Conclusions} is devoted to conclusion and discussions.

%%%%%%%%%%%%%%%%%%%%%%%%%%%%%%%%%%%%%%%%%%%%%%%%%%%%%%%%%%%%%%%%%%%%
\section{Large $N_f$ QCD}
\label{sec:large_Nf}

In QCD
with $N_f$ flavors of massless quarks,
the renormalization group equation (RGE) for the running coupling
$\alpha(\mu)  \ \Big(= \frac{\bar{g}^2(\mu)}{4 \pi}\,\Big)$ 
in the two-loop approximation is given by 
\begin{equation}
  \mu \frac{d}{d \mu} \alpha(\mu) 
  = -b \alpha^2(\mu) - c \alpha^3(\mu) ,
\label{eq:RGE_for_alpha}
\end{equation}
where
\begin{equation}
  b = \frac{1}{6 \pi} \left( 11 N_c - 2 N_f \right)\ ,\quad
  c = \frac{1}{24 \pi^2} \left( 34 N_c^2 - 10 N_c N_f
      - 3 \frac{N_c^2 - 1}{N_c} N_f \right) \ .
\end{equation}
{}From the above beta function we can easily see that,
when $b>0$ and $c<0$, i.e.,
$N_f$ takes a value in the range of 
$  N_f^* < N_f < \frac{11}{2} N_c $ 
($N_f^\ast \simeq 8.05$ for $N_c = 3$), 
the theory is asymptotically free and 
the beta function
has 
a zero, corresponding to
an infrared stable fixed 
point~\cite{Banks:nn,Appelquist:1996dq},
at
\begin{equation}
  \alpha_\ast = - \ \frac{\ b\ }{\ c\ } \ .
\label{eq:alpha_IR}
\end{equation}

Existence of the infrared fixed point 
implies that
the running coupling takes a finite value 
even in the low energy region.
Actually,
the solution of the two loop
RGE in Eq.~(\ref{eq:RGE_for_alpha}) can be explicitly 
written~\cite{Gardi,ExplicitSolution} 
in all the energy region
as 
\begin{equation}
  \alpha(\mu) = \alpha_\ast \left[\ W(\mu^{\, b \alpha_\ast} 
  / e \Lambda^{ b \alpha_\ast} ) + 1\ \right]^{-1},
\label{Lan W}
\end{equation}
where $W(x) = F^{-1}(x)$ with $F(x) = x e^x$ is the Lambert $W$
function, and 
$\Lambda$ is a renormalization group invariant
scale defined by~\cite{Appelquist:1996dq}
\begin{equation}
  \Lambda \ \equiv\  \mu \ \exp \left[ - \frac{1}{b\  \alpha_\ast}
         \log \left( \frac{\alpha_\ast - \alpha(\mu)}{\alpha(\mu)}
         \right)
        - \frac{1}{b\  \alpha(\mu)}
         \right] .
\label{eq:Lambda}
\end{equation}

\begin{figure}
  \begin{center}
    \includegraphics[height=6cm]{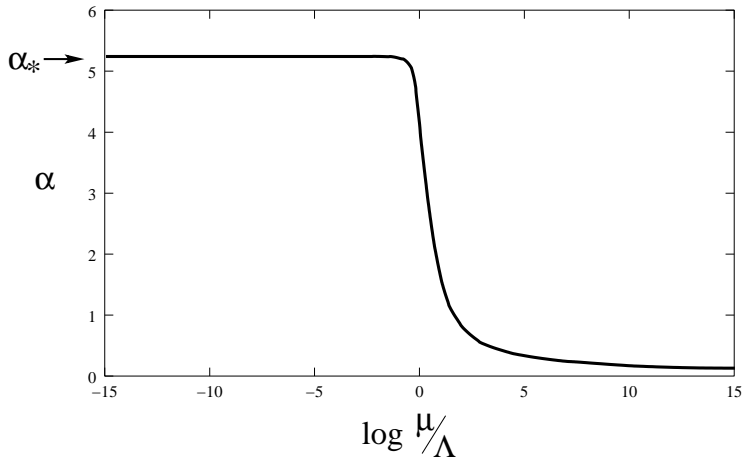}
  \end{center}
\caption{
Two-loop
running coupling of the large $N_f$ QCD for $N_f = 9$.
}
\label{fig:step}
\end{figure}

We note that, in the present analysis, we fix 
$\Lambda$ to compare the theories with a different number of flavors, 
and that we have no adjustable parameters in the
running coupling in Eq.~(\ref{Lan W}), accordingly.\footnote{ 
Note that we do not have to specify the value of $\Lambda$, since in this
paper we calculate only dimensionless quantities, $S$ parameter and 
$\Delta m_\pi^2/f_\pi^2$, which are independent of the actual value of $\Lambda$ in
unit of, say GeV. When we apply the dynamics to the model building for the electroweak
symmetry breaking, we should specify the value of $f_\pi$ (of order of the weak scale 250 GeV but 
actually model-dependent) and then determine the scale of
$\Lambda$ as a function of $N_c(=3)$ and $N_f$ as well as  $f_\pi$ (See Fig.\ref{fig:fpi_in_largeNf}).
} 
We show 
an example of $\alpha(\mu)$ for $N_f = 9$ 
in Fig.~\ref{fig:step}.
The fact that the running coupling is expressed by a
certain function as in Eq.~(\ref{Lan W}) implies that,
in the case of the large $N_f$ QCD,
{\it we do not need to introduce any infrared regularizations}
such as those adopted in Ref.~\cite{Harada:2004qn}, or in 
Refs.~\cite{Miransky:vj}
for studying real-life QCD with small $N_f$
in which
the infrared regularization parameter must be chosen 
in such a way that
the running coupling in the infrared region becomes larger
than the critical value $\alpha_{\rm cr} = \pi/4$
for realizing the dynamical chiral symmetry 
breaking. 
The running coupling in the large $N_f$ QCD takes a certain
value in the infrared region for given $N_f$, so that
we can definitely determine, within the framework of the SD equation,
whether or not the dynamical chiral symmetry breaking 
of 
$SU(N_f)_{\rm L}\times SU(N_f)_{\rm R} \rightarrow SU(N_f)_{\rm V}$
is realized.
Actually, the value of 
$\alpha_\ast$ decreases 
monotonically with increasing $N_f$,
and
the chiral symmetry restores when $N_f$ 
becomes large enough.
In Refs.~\cite{Appelquist:1996dq}, 
it was shown that the phase transition occurs at 
 $N_f^{\rm crit} \simeq  11.9$ for $N_c = 3$
(corresponding to $\alpha_\ast = \alpha_{\rm cr} = \pi/4$).
However, the value of $N_f^{\rm crit}$ itself
should not be taken seriously, since $\alpha_{\rm cr} = \pi/4$ to be equated to
$\alpha_\ast$ is already so
large as to invalidate the perturbation which determined the value 
of $\alpha_\ast$. There is also some ambiguity of the ladder approximation which
determined the value  $\alpha_{\rm cr} = \pi/4$ itself.
In fact lattice simulations~\cite{lattice} suggest smaller value $6<N_f^{\rm crit}<7$ and
other approaches also suggest different values~\cite{OZ,VS,Harada:1999zj,Gies:2005as}.

%%%%%%%%%%%%%%%%%%%%%%%%%%%%%%%%%%%%%%%%%%%%%%%%%%%%%%%%%%%%%%%%%%%%%%%
%%%%%%%%%%%%%%%%%%%%%%%%%%%%%%%%%%%%%%%%%%%%%%%%%%%%%%%%%%%%%%%%%%%%%%%

\section{Expressions of $f_\pi$, ${\hat S}$ and $\Delta m_\pi^2$ in
 terms of current correlators}
\label{sec:Spectral}
 
 In this section we introduce sum rules for expressing the order
parameter $f_\pi$, the parameter ${\hat S}$ which is the contribution
to the $S$ parameter from one weak fermionic doublet, and the square
mass of the pseudo NG $\Delta m_\pi^2$ in terms of $\Pi_{VV} -
\Pi_{AA}$ in the space-like momentum region.
 
 Let us begin with
 introducing the vector and the axial-vector currents as 
 \begin{eqnarray}
   V_\mu^a(x) &=& \bar\psi(x) T^a \gamma_\mu \psi(x)\, ,
 \nonumber\\
   A_\mu^a(x) &=& \bar\psi(x) T^a \gamma_\mu \gamma_5
   \psi(x)\, ,
  \label{def:currents}
 \end{eqnarray}
 where 
 $T^a$ ($a=1,2,\ldots,N_f^2-1$) is the generator of 
$SU(N_f)$ normalized as 
 $\mbox{tr}(T^a T^b) = \frac{1}{2}
 \delta^{ab}$.
As we discussed in the previous section, for $N_f < N_f^{\rm crit}$
 there occurs the spontaneous symmetry breaking of
$SU(N_f)_{\rm L}\times SU(N_f)_{\rm R} \rightarrow SU(N_f)_{\rm V}$.
As a result, the massless NG boson $\pi$ appears, whose decay constant
 $f_\pi$ is defined as
 \begin{equation}
   \langle 0 \vert A_\mu^a (0) \vert \pi^b(q) \rangle 
   = i q_\mu f_\pi \delta^{ab}\, ,
 \end{equation}
where $a,b=1,2,\ldots,N_f^2-1$. 
As is well-known as 
the first Weinberg sum rule~\cite{Weinberg:1967kj}, 
the above decay constant is related the difference of the
vector current correlator $\Pi_{VV}$ and 
the axial-vector current correlator $\Pi_{AA}$ as
\begin{equation}
   f_\pi^2 = \Pi_{VV}(0) - \Pi_{AA}(0)\, ,
 \label{eq:fpi}
\end{equation}
where the correlators are defined from the currents
in Eq.~(\ref{def:currents}) as 
 \begin{eqnarray}
   \delta^{ab} \left( \frac{q_\mu q_\nu}{q^2} - g_{\mu \nu} \right) 
 \Pi_{JJ}(q^2) &=& i 
   \int d^4x\ e^{i q x}\ \langle 0 \vert T J_\mu^a(x) J_\nu^b(0) 
   \vert 0 \rangle\, , \\ 
 & & \hspace{2cm} 
     ( J_\mu^a(x) = V_\mu^a(x), A_\mu^a(x) )\, . \nonumber
 \end{eqnarray}

When the large $N_f$ QCD is considered as the underlying theory
for describing the Higgs sector of the standard model as in 
the technicolor-like scenario \cite{TC, techni},
it is important to study the $S$ parameter~\cite{PeskinTakeuchi}
which gets a strong constraint from the electroweak precision 
tests~\cite{Eidelman:2004wy}.
Since entire contributions to the $S$ parameter depend on the
detail of the model structure, we concentrate on 
calculating the contribution 
from one $SU(2)_{\rm L}$ doublet with the hypercharge $Y=1/6$ and
two singlets with $Y=2/3$ and $Y=-1/3$.
This parameter $\hat{S}$ is related to the above correlators as
\begin{equation}
   {\hat S} = \left.
   - 4 \pi \frac{d}{d Q^2} 
   \left[ \Pi_{VV}(Q^2) - \Pi_{AA}(Q^2)
   \right] \right\vert_{Q^2 = 0}\, ,
 \label{eq:S_parameter}
\end{equation}
which is nothing but the DMO sum
rule~\cite{Das:1967ek} or    
often called the ``zeroth Weinberg
sum rule''.

Mass of the pseudo NG boson is also an important quantity
in the technicolor-like scenario.
In the present analysis,
we consider the situation that 
$SU(N_f)_L \otimes SU(N_f)_R$ 
chiral symmetry is explicitly broken by the $U(1)$ gauge symmetry,
the generator of which is given as in Eq.~(\ref{def:Q}),
and calculate the 
the mass of the pseudo NG boson associated with the
generator given in Eq.~(\ref{eq:pNGB}).
The squared mass of the above
pseudo NG boson $\Delta m_\pi^2$ is related to the 
above correlators as~\cite{Yamawaki:1982kt} 
 \begin{equation}
   \Delta m_\pi^2 
   = \frac{3 \alpha_{em}}{4 \pi f_\pi^2} 
     \int_{0}^{\infty} d Q^2 \left[ \Pi_{VV}(Q^2) -
   \Pi_{AA}(Q^2) \right]\, .
 \label{eq:mass_diff}
 \end{equation}
where
$\alpha_{em} (= \frac{e^2}{4 \pi} = \frac{1}{137})$.
Equation~(\ref{eq:mass_diff}) was first derived by 
Das et al.~\cite{Das:it} in the form written in terms of the spectral function instead of
the current correlators
and may be called the
DGMLY sum rule or the 
``third Weinberg sum rule'' .

As was studied in, e.g., Ref.~\cite{Peskin-Preskil},
$\Delta m_\pi^2$ is interesting quantity to study because it is 
 related to the structure of the vacuum.
 When we switch off the above gauge interaction,
we have $N_f^2-1$ massless NG bosons
associated with the spontaneous
 breaking of 
 $SU(N_f)_{\rm L} \otimes SU(N_f)_{\rm R}$ chiral symmetry 
 down to $SU(N_f)_{\rm V}$ symmetry.
 The existence of the gauge interaction explicitly 
 breaks the $SU(N_f)_{\rm L} \otimes SU(N_f)_{\rm R}$ chiral symmetry,
which makes some of NG bosons (say $\pi^+$) be massive while
others (say $\pi^0$) remain massless.
The interesting point here is whether 
 $\Delta m_\pi^2$ becomes
 positive or negative, 
which is called ``vacuum alignment problem''~\cite{Peskin-Preskil}.
 Negative $\Delta m_\pi^2$ means that fluctuation of $\pi^+$ field 
 around $\langle\pi^+\rangle = 0$ is unstable and 
 the vacuum with 
 $\langle\pi^+\rangle = 0$ is not a true vacuum.
 If this is the case, $\pi^+$ has non-zero vacuum expectation value 
 and $U(1)$ gauge symmetry is broken.
In the case of the real-life ($N_f=3$) QCD,
we have shown~\cite{Harada:2004qn} that positive 
 $\Delta m_\pi^2$ is successfully reproduced in the framework of 
 the SD equation and the IBS equation with the improved ladder 
 approximation.
In our best knowledge, however, no one has explicitly shown 
whether positive 
 $\Delta m_\pi^2$ is also realized in the case of the large $N_f$ 
 QCD, especially near the critical point, so
our calculation is the first one which explicitly investigates 
the stability of the vacuum in the large $N_f$ QCD.

 %%%%%%%%%%%%%%%%%%%%%%%%%%%%%%%%%%%%%%%%%%%%%%%%%%%%%%%%%%%%%%%%%%%%
 \section{Current correlators from BS amplitudes}
 \label{sec:correlator-BS}
 
 In the previous section we have written down the sum rules for
$f_\pi$, $\hat{S}$
and $\Delta m_\pi^2$ in terms of the current correlators.
 In this section, we show how the current correlators 
 are obtained from BS amplitudes which will be
 calculated from the IBS equation.

For deriving several properties of hadrons as boundstates, 
we often need to 
 perform calculations in the time-like momentum region.
 However, it is difficult to solve the BS equation and the SD equation 
 in the time-like region 
 since we have to carry out the analytic continuation of the running
 coupling from the space-like region to the time-like 
region.~\footnote{
 In general, one-loop running coupling of QCD is not analytic, 
 so that several models for running coupling are introduced
 to solve the SD and BS equations in the time-like region.
 On the other hand, the running coupling at two-loop level
 becomes fully analytic 
 near the phase transition point in the large $N_f$ QCD.
 In Ref.~\cite{Harada:2003dc}, we approximated the analytic 
 running coupling by a simple function
 $\frac{\bar{g}^2(x+y)}{4\pi} 
  = \alpha_\ast\ \theta(\Lambda^2 - (x+y) )$,
 and studied the critical behaviors of $f_\pi$ as well as masses and
 decay constants of scalar, vector, and axial-vector mesons.
 We note that, in the present analysis, we do not use the 
 approximate form, but use the fully analytic two-loop running 
 coupling.
}
In the present analysis, on the other hand,
we need the BS amplitude only
 in the space-like region in order to calculate the current
 correlators appearing in the sum rules
 for $\Delta m_\pi^2$, $f_\pi$ and the QCD $S$ parameter.
 
Now, the  
 BS amplitudes $\chi^{(J)}$ ($J=V$, $A$) 
 are defined in terms of the three-point vertex function 
 as follows:
 \begin{equation}
   \delta_i^j \left( T^a \right)_{f}^{f'} 
   \int \frac{d^4 p}{(2 \pi)^4} e^{- i p r} \chi^{(J)}_{\alpha
   \beta}(p;q,\epsilon) = 
   \epsilon^\mu \int d^4 x e^{i q x} 
   \langle 0 \vert T\ \psi_{\alpha i f}(r/2)\ 
   \bar\psi_\beta^{j f'}(-r/2)\ J_\mu^a(x)\ 
   \vert 0 \rangle,
 \label{eq:three-point}
 \end{equation}
 where $q^\mu$ is the total momentum of the fermion and
 the anti-fermion, and $p^\mu$ is the relative one.
 $\epsilon^\mu$ is the polarization vector defined by 
 $\epsilon \cdot q = 0$, $\epsilon \cdot \epsilon = -1$, 
 and $(f, f'), (i, j), (\alpha, \beta)$ are flavor, color and 
 spinor indices, respectively.
 Closing the fermion legs of 
 the above three-point vertex function 
 and taking the limit $r \rightarrow 0$, 
 we can express the current correlator in terms of the 
 BS amplitude as follows:
 \begin{equation}
   \Pi_{JJ}(q^2) = \frac{1}{3} \sum_{\epsilon} \int \frac{d^4 p}{i (2
   \pi)^4} \frac{N_c}{2} \mbox{tr} \left[ 
   \left(\epsilon \cdot G^{(J)}\right) 
   \chi^{(J)}(p;q,\epsilon) \right]
 \label{eq:Pi_JJ}
 \end{equation}
 where 
 \begin{equation}
   G_\mu^{(V)} = \gamma_\mu,\ \ \ \ G_\mu^{(A)} = \gamma_\mu \gamma_5,
 \end{equation}
 and $N_c = 3$ is the number of colors.
 In the above expression we averaged over the polarizations 
 so that $\Pi_{JJ}(q^2)$ does not depend on the polarization.

 We expand the BS amplitude  
 $\chi_{\alpha \beta}^{(J)}(p;q,\epsilon)$  
 in terms of the bispinor bases 
 $\Gamma^{(J)}_i$ and the invariant amplitudes $\chi^{(J)}_i$ 
 as 
 \begin{equation}
   \left[ \chi^{(J)}(p;q,\epsilon) \right]_{\alpha \beta} 
   =\ \sum_{i=1}^{8} \left[ \Gamma_i^{(J)}(p;\hat{q},\epsilon)
   \right]_{\alpha \beta} \chi^{(J)}_{i} (p;q) ,
 \label{eq:chi-expanded}
 \end{equation}
 where $\hat{q}_\mu = q_\mu/\sqrt{Q^2}$ with $Q^2=-q^2$.
 The bispinor bases can be chosen in a way that they have the same
 properties of 
 spin, parity 
 and charge conjugation as the corresponding
 current $J_\mu^a(x)$ has.
 We adopt the following bispinor bases for the vector vertex:
 \begin{eqnarray}
 & &  \Gamma^{(V)}_1 = \fsl{\epsilon} ,\ \ 
      \Gamma^{(V)}_2 = \frac{1}{2} [\fsl{\epsilon},\fsl{p}] 
                            (p \cdot \hat q) ,\ \ 
      \Gamma^{(V)}_3 = \frac{1}{2} [\fsl{\epsilon},\fsl{\hat q}] ,\ \ 
      \Gamma^{(V)}_4 = \frac{1}{3!}[\fsl{\epsilon},\fsl{p},\fsl{\hat q}]
      \\  
 & &  \Gamma^{(V)}_5 = (\epsilon \cdot p) ,\ \ 
      \Gamma^{(V)}_6 = \fsl{p} (\epsilon \cdot p) ,\ \ 
      \Gamma^{(V)}_7 = \fsl{\hat q}(p \cdot \hat q) 
            (\epsilon \cdot p) ,\ \ 
      \Gamma^{(V)}_8 = \frac{1}{2} [\fsl{p},\fsl{\hat q}]
            (\epsilon \cdot p) ,
 \nonumber
 \end{eqnarray}
 where $[a,b,c] \equiv a[b,c] + b[c,a] + c[a,b]$.
 For the axial-vector vertex we use 
 \begin{eqnarray}
   \Gamma^{(A)}_1 &=& \fsl{\epsilon}\ \gamma_5 ,\ \ \ 
      \Gamma^{(A)}_2 = \frac{1}{2} [\fsl{\epsilon},\fsl{p}] 
                            \gamma_5  ,\ \ \ 
      \Gamma^{(A)}_3 = \frac{1}{2} [\fsl{\epsilon},\fsl{\hat q}]\ 
              (p \cdot \hat q)
                 \ \gamma_5  ,\nonumber\\
      \Gamma^{(A)}_4 &=& \frac{1}{3!}[\fsl{\epsilon},\fsl{p},
              \fsl{\hat q}]
      \ \gamma_5 ,\ \ \ 
      \Gamma^{(A)}_5 = (\epsilon \cdot p)\ (p \cdot \hat q)\ 
        \gamma_5  ,\ \ \ 
      \Gamma^{(A)}_6 = \fsl{p} (\epsilon \cdot p)\ \gamma_5  ,
      \nonumber\\
      \Gamma^{(A)}_7 &=& \fsl{\hat q}\ (\epsilon \cdot p)\ 
             (p \cdot \hat q)
                 \ \gamma_5  ,\ \ \ 
      \Gamma^{(A)}_8 = \frac{1}{2} [\fsl{p},\fsl{\hat q}]
             (\epsilon \cdot p)\
      (p \cdot \hat q)\ \gamma_5.
 \end{eqnarray}
 {}From the above choice of the bases, we can easily show that all the
 invariant amplitudes $\chi_i^{(J)}$ are the even functions of 
 $(p\cdot \hat{q})$ using the charge conjugation property of the
 current.

 In the present analysis we fix the frame of reference in such a way 
 that only the zero component of the total momentum $q^\mu$ becomes 
 non-zero.
 Furthermore, we study the case where $q^\mu$ is in the space-like 
 region.
 Then, it is convenient to parameterize the total momentum $q^\mu$ as
 \begin{equation}
   q^\mu = (i Q , 0 , 0 , 0 ) .
 \end{equation}
 For the relative momentum $p^\mu$, we perform 
 the Wick rotation, and parameterize it by the real 
 variables $u$ and $x$ as
 \begin{equation}
   p \cdot q = - Q\  u \ ,\ \  p^2 = - u^2 - x^2 .
 \end{equation}
 Consequently, the invariant amplitudes $\chi^{(J)}_i$
 become functions 
 in $u$ and $x$:
 \begin{equation}
   \chi_i^{(J)} = \chi_i^{(J)}(u,x;Q) .
 \end{equation}
 {}From the charge conjugation properties for the  
 BS amplitude $\chi^{(J)}$ 
 and the bispinor bases defined above, 
 the invariant amplitudes $\chi_i^{(J)}(u,x)$ are shown to 
 satisfy 
 \begin{equation}
   \chi_i^{(J)}(u,x;Q) = \chi_i^{(J)}(-u,x;Q)\ .
 \label{eq:even-property}
 \end{equation}
 Using this property of the invariant amplitudes, 
 we rewrite Eq.~(\ref{eq:Pi_JJ}) as 
 \begin{equation}
   \Pi_{VV}(Q^2) \ =\  \frac{N_c}{\pi^3} \int_0^\infty du 
   \int_0^\infty dx\ x^2 \left[ - \chi^{(V)}_1(u,x;Q) 
   + \frac{x^2}{3} \chi^{(V)}_6(u,x;Q) 
   \ \right],
 \label{eq:PiVV}
 \end{equation}
 \begin{equation}
   \Pi_{AA}(Q^2) \ =\  \frac{N_c}{\pi^3} \int_0^\infty du 
   \int_0^\infty dx\ x^2 \left[\ \ \chi^{(A)}_1(u,x;Q) 
   - \frac{x^2}{3} \chi^{(A)}_6(u,x;Q) 
   \ \right].
 \label{eq:PiAA}
 \end{equation}
 Here, we used the expanded form of the BS amplitude shown in 
 Eq.~(\ref{eq:chi-expanded}) and carried out the three dimensional 
 angle integration.
 
 From Eqs.~(\ref{eq:PiVV}) and (\ref{eq:PiAA}), 
 the quantity $\Pi_{VV} - \Pi_{AA}$ is expressed as
 \begin{eqnarray}
   \Pi_{VV} - \Pi_{AA} &=&  
   \frac{1}{3} \sum_{\epsilon} \int \frac{d^4 p}{i (2
   \pi)^4} \frac{N_c}{2} \mbox{tr} \left[ 
   \fsl{\epsilon} \chi^{(J)}(p;q,\epsilon) - 
   \fsl{\epsilon} \gamma_5 \chi^{(A)}(p;q,\epsilon) 
   \right],\nonumber\\
   &=& 
   \frac{N_c}{\pi^3} \int_0^\infty du 
   \int_0^\infty dx\ x^2 \Bigg[ - \left( \chi^{(V)}_1(u,x;Q) 
   + \chi^{(A)}_1(u,x;Q) \right) \nonumber\\
   & &\ \ \ \ \ \ \ \ \ \ \ \ \ \ \ \ \ \ \ \ \ 
      + \frac{x^2}{3} \left( \chi^{(V)}_6(u,x;Q) 
   + \chi^{(A)}_6(u,x;Q) \right) 
   \ \Bigg].
 \end{eqnarray}
 We note that, although either $\Pi_{VV}$ or $\Pi_{AA}$ is
 logarithmically divergent quantity, the difference
 $\Pi_{VV} - \Pi_{AA}$ becomes 
 finite due to the cancellation of the divergence ensured by
 the chiral symmetry.

 %%%%%%%%%%%%%%%%%%%%%%%%%%%%%%%%%%%%%%%%%%%%%%%%%%%%%%%%%%%%%%%%%%%%
 %%%%%%%%%%%%%%%%%%%%%%%%%%%%%%%%%%%%%%%%%%%%%%%%%%%%%%%%%%%%%%%%%%%%
 \section{Inhomogeneous Bethe-Salpeter equation}
 \label{sec:IBS}
 
 In this section we introduce the inhomogeneous Bethe-Salpeter
 (IBS) equation 
 from which we calculate the BS amplitude 
 defined in the previous section.
 
\begin{figure}
   \begin{center}
     \includegraphics[height=3cm]{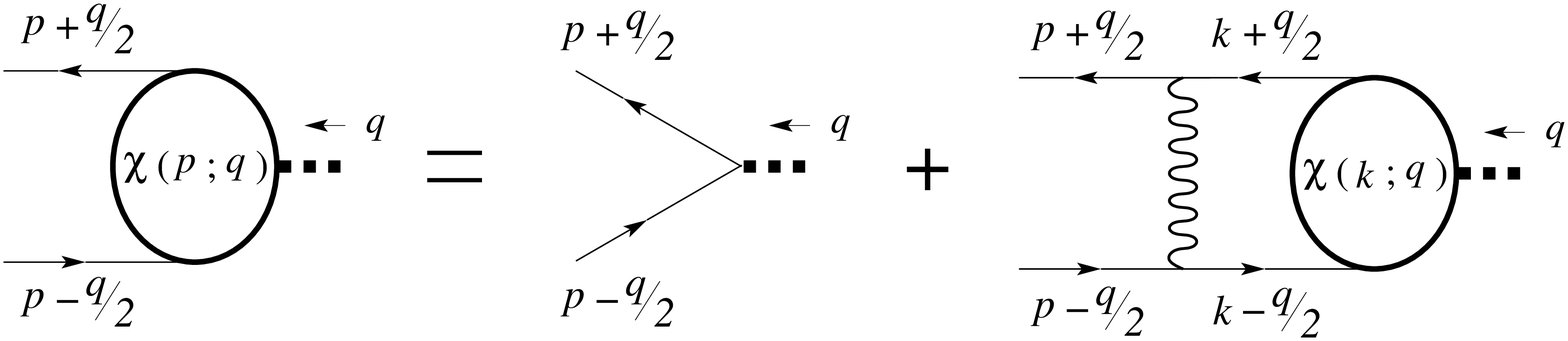}
   \end{center}
 \caption{A graphical expression of the IBS equation in the 
 (improved) ladder approximation.}
 \label{fig:IBSeq}
 \end{figure}

 The IBS equation is the self-consistent equation for 
 the BS amplitude $\chi^{(J)}$, and it is expressed as 
 (see Fig.~\ref{fig:IBSeq} for graphical expression)
 \begin{equation}
   T(p;q)\ \chi^{(J)}(p;q,\epsilon) \ \ =\ \ 
   \epsilon \cdot G^{(J)}\ 
   +\ K(p;k) \ast \chi^{(J)}(k;q,\epsilon) .
 \label{eq:IBSeq}
 \end{equation}
 The kinetic part $T$ is given by 
 \begin{equation}
   T(p;q) =   S_F^{-1}(p + q/2) \otimes  S_F^{-1}(p - q/2) \ ,
 \label{def T}
 \end{equation}
 where $S_F$ is the full fermion propagator 
 $i S_F^{-1}(p) = \fsl{p} - \Sigma(p) $. 
 (Note that wave function renormalization factor $A(p)$ becomes unity 
 when we adopt the Landau gauge.)
 The BS kernel $K$ in the improved ladder approximation 
 is expressed as
 \begin{equation}
   K(p;k) \ =\    \frac{N_c^2 - 1}{2 N_c} \ 
          \frac{\bar{g}^2(p,k)}{- (p-k)^{2}}\ 
          \left( g_{\mu\nu} - \frac{(p-k)_\mu (p-k)_\nu}{(p-k)^2}
          \right) \cdot \gamma^\mu \otimes \gamma^\nu ,
 \end{equation} 
 where $\bar{g}(p,k)$ is the running coupling of QCD whose explicit 
 form will be shown later.
 In the above expressions we used
 the tensor product notation
 \begin{equation}
   (A \otimes B) \,\chi  =  A\, \chi\, B \ ,
 \end{equation}
 and the inner product notation 
 \begin{equation}     K(p;k) \ast \chi^{(J)}(k;q,\epsilon) =   
   \int \frac{d^4 k}{i(2\pi)^4}\  K(p,k)\  \chi(k;q) \ .
 \end{equation}
 
 The mass function of the quark propagator is obtained from 
 the SD equation (see appendix \ref{app:SD}) : 
 \begin{equation}
   \Sigma(p) = K(p,k) \ast i S_F(p).
 \end{equation}
 It should be stressed that we must
 use the same kernel $K(p,k)$ as that used in the 
 IBS equation for consistency with the chiral 
 symmetry~\cite{ Kugo:1992pr,Bando-Harada-Kugo}.
Numerical method for solving the SD equation and the IBS equation 
are shown in appendix \ref{app:num_SD} and appendix \ref{app:num_IBS}, 
respectively.

%%%%%%%%%%%%%%%%%%%%%%%%%%%%%%%%%%%%%%%%%%%%%%%%%%%%%%%%%%%%%%%%%%%%%%
%%%%%%%%%%%%%%%%%%%%%%%%%%%%%%%%%%%%%%%%%%%%%%%%%%%%%%%%%%%%%%%%%%%%%%
\section{Numerical results}
\label{sec:results}

In this section, we show the results of the calculations for $f_\pi$, 
${\hat S}$, and $\Delta m_\pi^2$ in the large $N_f$ QCD near the chiral
phase transition point. 
The procedure of the calculation is as follows.
First, we solve the IBS equation and the SD equation simultaneously, 
and obtain BS amplitude $\chi^{V, A}$.
(Details of numerical methods to
solve the SD and the IBS equations are shown in
appendix~\ref{app:num_SD} and appendix~\ref{app:num_IBS}.)
Then, by closing the fermion legs of BS amplitudes, we obtain current
correlators $\Pi_{VV}(Q^2)$ and $\Pi_{AA}(Q^2)$. 
(See section~\ref{sec:correlator-BS}.) 
Once we obtain $\Pi_{VV}(Q^2)$ and $\Pi_{AA}(Q^2)$, we 
calculate $f_\pi^2$, ${\hat S}$, and $\Delta m_\pi^2$ from the
relations  
in Eqs.~(\ref{eq:fpi}), (\ref{eq:S_parameter}) and
(\ref{eq:mass_diff}).

%%%%%%%%%%%%%%%%%%%%%%%%%%%%%%%%%%%%%%%%%%%%%%%%%%%%%%%%%%%%%%%%%%%%%%
\subsection{Critical behavior of $f_\pi$ in the large $N_f$ QCD}
\label{sec:fpi_result}

\begin{figure}
  \begin{center}
    \includegraphics[height=8.5cm]{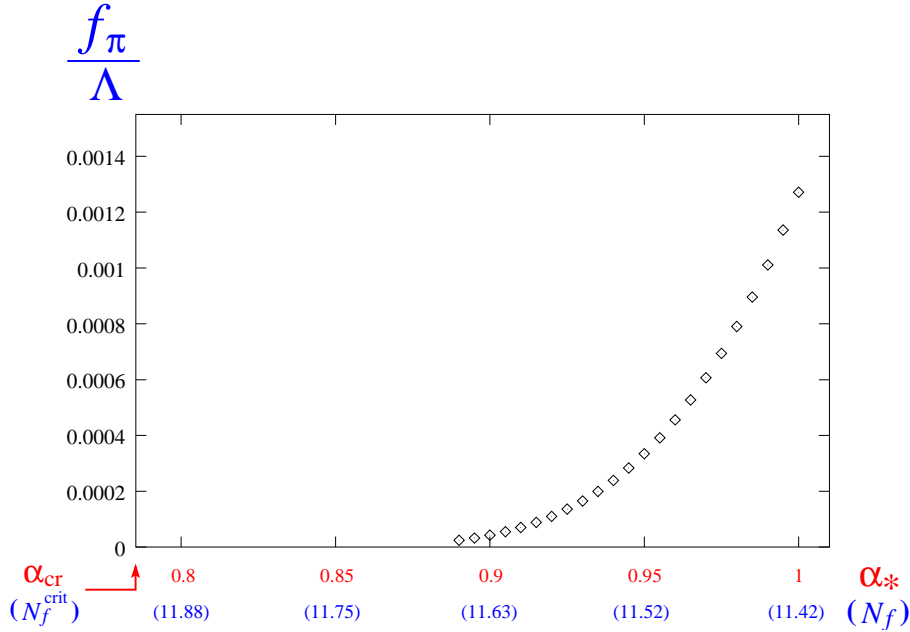}
  \end{center}
\caption{Values of $f_\pi$ calculated from Eq.~(\ref{eq:fpi}) for
 several values of $\alpha_\ast$ in the range of
$\alpha_\ast \in [0.89:1]$ (indicated by $\Diamond$).
Values of $N_f$ corresponding to each $\alpha_\ast$ are shown 
in parenthesis.}
\label{fig:fpi_in_largeNf}
\end{figure}

In Fig.~\ref{fig:fpi_in_largeNf}, we plot $f_\pi$ for
several values of $\alpha_\ast$ in the range of
$\alpha_\ast \in [0.89:1]$.
Note that we use $\alpha_\ast$ instead of $N_f$ as an input
parameter, because once we choose a value of $N_f$, the value of
$\alpha_\ast$ is uniquely determined from Eq.~(\ref{eq:alpha_IR}). 
$\alpha_\ast = \alpha_{\rm cr}$ implies $N_f = 11.91$. 
In Fig.~\ref{fig:fpi_in_largeNf}, we have shown corresponding 
values of $N_f$ in the parenthesis. 
From this figure, we can see that $f_\pi$ goes to zero when we reduce
the value of $\alpha_\ast$ to its critical value $\alpha_{\rm cr} \simeq  
\pi/4$ (or increase the value of $N_f$ to its critical value 
$N_f^{\rm crit} \simeq 4 N_c$), 
which is consistent with the chiral phase transition described
in section~\ref{sec:large_Nf}.
Furthermore, the scaling behavior is identified as essential
singularity type, 
\begin{equation}
 f_\pi = p \,\Lambda \,\exp 
       \left(
            - \frac{q}{\sqrt{\frac{\alpha\ast}{\alpha_{\rm cr}}-1\ }}
       \right)
\label{fpilargenf}
\end{equation}
with
$p=15.1$ and 
$q=4.9 $ as the best fit. This is similar to the case with constant coupling with sharp cutoff
at $\Lambda$
%which was already confirmed in
~\cite{Harada:2003dc}:\footnote{
At first glance there is some numerical difference between the case with a 
constant coupling and that with a two-loop coupling, which is mainly due to the different
meanings of $\Lambda$ for these two cases.   Actually, such a superficial difference 
 also takes place for  $m=\Sigma(m^2)$ which is essentially proportional to $f_\pi$, 
with the ratios being the same in both cases.
}
\begin{equation}
  f_\pi = d \,\Lambda \,\exp 
       \left(
            - \frac{\pi}{\sqrt{\frac{\alpha\ast}{\alpha_{\rm cr}}-1\ }}
       \right)
   \ , \quad (d\simeq 1.5)\, .
\label{fpi}
\end{equation}

\subsection{Critical behavior of $\Delta m_{\pi}^2$ in the large $N_f$ QCD}
\label{sec:massdiff_result}

\begin{figure}
 \begin{center}
  \includegraphics[height=8.5cm]{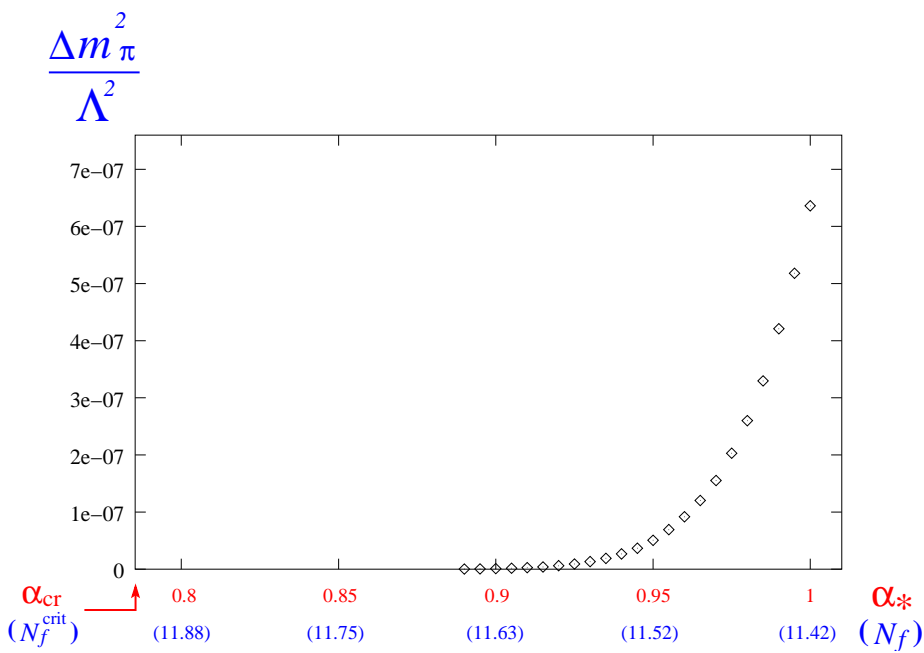}
 \end{center}
 \caption{Values of $\Delta m_{\pi}^2$ calculated from 
 Eq.~(\ref{eq:mass_diff}) for several values of $\alpha_\ast$ in the
 range of $\alpha_\ast \in [0.89:1]$ (indicated by $\Diamond$). 
 }
 \label{fig:massdiff_in_largeNf}
\end{figure}

In Fig.~\ref{fig:massdiff_in_largeNf}, we plot $\Delta m_{\pi}^2$
calculated for several values of $\alpha_\ast$ in the range of 
$\alpha_\ast \in [0.89:1]$.
In our numerical calculation of $\Delta m_\pi^2$, we introduced 
the numerical cutoff $\Lambda_{\rm Num}^2$: 
\begin{equation}
 \frac{\Pi_{V-A}(Q^2=\Lambda_{\rm Num}^2)}{\Pi_{V-A}(Q^2=0)} 
 = \frac{1}{50}
\label{eq:numerical_cutoff}
\end{equation} 
for the integration of 
$ \Pi_{V-A} (Q^2) \equiv \left[ \Pi_{VV}-\Pi_{AA}\right](Q^2)$ in Eq.~(\ref{eq:mass_diff}) 
as follows:
\begin{equation}
   \Delta m_\pi^2 
   = \frac{3 \alpha_{em}}{4 \pi f_\pi^2} 
     \int_{0}^{\Lambda_{\rm num}^2} d Q^2 \,
%\left[ \Pi_{VV}(Q^2) -  \Pi_{AA}(Q^2) \right]
\Pi_{V-A} (Q^2)
\, .
 \label{eq:mass_diff_num}
\end{equation}
%where $\Lambda_{\rm Num}^2$ is defined by
We shall discuss later inclusion of the contributions from $\Lambda_{\rm Num}^2 < Q^2< \Lambda^2$. 

\begin{figure}
 \begin{center}
  \includegraphics[height=8.5cm]{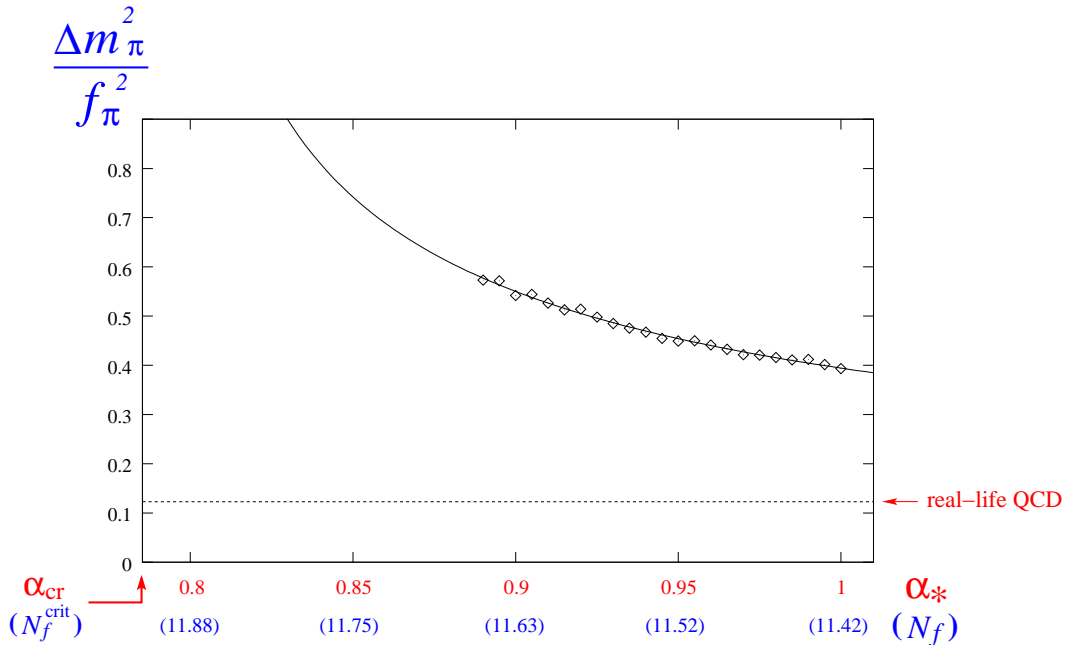}
  \end{center}
 \caption{Values of $\Delta m_{\pi}^2/f_\pi^2$ for several values of
  $\alpha_\ast$ in the 
 range of $\alpha_\ast \in [0.89:1]$ (indicated by $\Diamond$).  
  Solid line shows the fitting function $I(\alpha_\ast)$ 
  given in Eq.(\ref{inversesqrtfit}).
Value of $\Delta m_{\pi}^2/f_\pi^2$ of the real-life QCD calculated in 
 Ref.~\cite{Harada:2004qn} is also shown in the figure.
Here, we used the data 
with the infrared cutoff parameter of the running coupling 
$t_F=0.04$.
 }
 \label{fig:massdiff_over_fpi2}
\end{figure}

From this figure, we see that $\Delta m_{\pi}^2$ vanishes to zero when
approaching the chiral phase transition point from the broken phase.
To see the scaling behavior of $\Delta m_{\pi}^2$ near the phase
transition point, we plot the values of $\Delta m_{\pi}^2/f_\pi^2$ in 
Fig.~\ref{fig:massdiff_over_fpi2} for several values of
$\alpha_\ast$ in the range of $\alpha_\ast \in [0.89:1]$.
From this figure, we can see that $\Delta m_{\pi}^2/f_\pi^2$ in the
large $N_f$ QCD near the phase transition point gradually increases
\begin{equation}
 \frac{\Delta m_{\pi}^2}{f_\pi^2} 
= 0.4 \quad {\rm to} \quad 0.6,
\label{eq:mass_diff_over_fpi2}
\end{equation}
in the range of $\alpha_\ast \in [0.89:1]$ investigated
here.  

The most outstanding feature of the value in Eq.~(\ref{eq:mass_diff_over_fpi2})
is that it is rather large compared with 
that of the real-life QCD~\cite{Harada:2004qn},
\begin{equation}
 \frac{\Delta m_{\pi}^2}{f_\pi^2}\Big |_{\rm real-life QCD} \simeq 0.123 \, ,
\end{equation}
calculated in the same method with the numerical cutoff $\Lambda_{\rm Num}$ defined as
Eq.~(\ref{eq:numerical_cutoff})
(The experimental value:   
$\Delta m_\pi^2/f_\pi^2 =0.148\pm 0.001$).
%This is a result unexpected in the usual discussions of the walking/conformal (scale-invariant) technicolor~\cite{Yamawaki:1985zg} where
%only discussed was the enhanced ${\rm mass}^2$ of pseudo NG boson due to the ETC interactions but not the
%${\rm mass}^2$ from the radiative interactions analogous to $\Delta m_\pi^2/f_\pi^2$ (apart from a
%simple scaling factors of $(1/N_D)(3/N_{TC})$, where $N_D=N_f/2$ and $N_{TC}$ are the number of 
%weak-doublet technifermions and the $SU(N_{TC})$-technicolors, respectively).
We shall argue that this  enhancement %factor of roughly four %is also due to 
reflects the large anomalous 
dimension $\gamma_m \simeq 1$
characteristic to the walking/conformal (scale-invariant) technicolor~\cite{Yamawaki:1985zg}. 
%When it is applied to the mass of $P^\pm$
%in the typical one-family model (Farhi-Susskind model), which corresponds to a large
%$N_f$ QCD with $N_f=8$,
%the prediction would be 
%changed from the standard estimate of 100 GeV region~\cite{techni} to 200 GeV region,
%which could be of phenomenological relevance.

\subsection{Inverse square root scaling of $\Delta m_\pi^2$}
\label{sec:inversesqrtsc}
\begin{figure}
 \begin{center}
  \includegraphics[height=8.5cm]{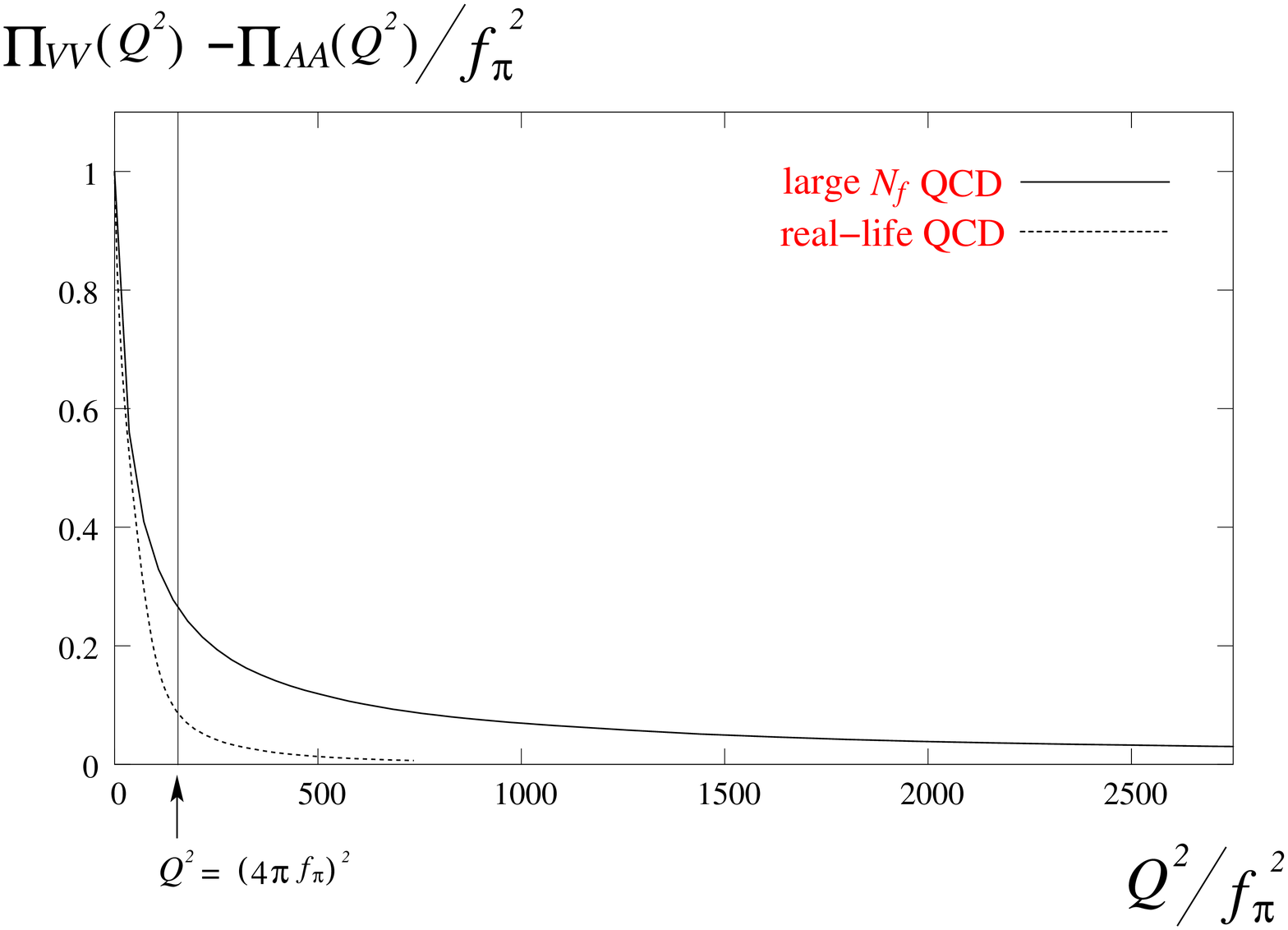}
 \end{center}
 \caption{
Comparison of $\Pi_{V=A}(Q^2)%-\Pi_{AA}(Q^2)
$ between the large $N_f$ 
QCD and the real-life QCD. Solid line indicates 
$\Pi_{V-A}(Q^2)
%-\Pi_{AA}(Q^2)
$ in the large $N_f$ QCD for
$\alpha_\ast=0.90$, while dotted line indicates
$\Pi_{V-A}(Q^2)
%-\Pi_{AA}(Q^2)
$ in the real-life QCD ($N_f=3$).
In this figure, both horizontal and vertical axes are normalized by the respective
value of 
$f_\pi^2$. $\Pi_{V-A}(Q^2)
%-\Pi_{AA}(Q^2)
$ in the real-life QCD was 
calculated in Ref.~\cite{Harada:2004qn}. Here, we used the data 
with the infrared cutoff parameter of the running coupling $t_F=0.04$,
which reproduce the QCD $S$ parameter and $\Delta m_\pi^2$ in good 
agreement with experiments ~\cite{Harada:2004qn}.
}
 \label{fig:largeNf_vs_realQCD.eps}
\end{figure}

To see the cause of this enhancement of  $\Delta m_\pi^2/f_\pi^2$ 
let us compare the behavior of
$\Pi_{V-A}(Q^2)
% - \Pi_{AA}(Q^2))
/f_\pi^2$ 
versus $Q^2/f_\pi^2$ of 
the large $N_f$ QCD with that of the real-life QCD, 
see Fig.~\ref{fig:largeNf_vs_realQCD.eps}. 
It is clear from Fig.~\ref{fig:largeNf_vs_realQCD.eps} that
$\Pi_{V-A}(Q^2)
%-\Pi_{AA}(Q^2))]
/f_\pi^2$ in the large $N_f$ QCD
is more slowly damping in high $Q^2$ than that in the real-life QCD,
which consequently yields bigger area integral in 
Fig.~\ref{fig:largeNf_vs_realQCD.eps} than that of the real-life QCD.
This is the cause of the enhancement of $\Delta m_\pi^2/f_\pi^2$. 

To understand the slow damping it is convenient to divide the integral region 
of the DGMLY sum rule (``third Weinberg sum rule''),
Eq.~(\ref{eq:mass_diff}),
into high and low energy regions as
\begin{eqnarray}
 \Delta m_\pi^2 
  &=& \frac{3 \alpha_{em}}{4 \pi f_\pi^2} 
  \left(
  \int_{0}^{\Lambda_\chi^2} d Q^2 + 
  \int_{\Lambda_\chi^2}^{\infty} d Q^2
  \right)\,
  %\left[ 
\Pi_{V-A}(Q^2)
% - \Pi_{AA}(Q^2) \right] 
\nonumber\\
 &\equiv& 
  \Delta m_{\pi ({\rm IR})}^2 + \Delta m_{\pi ({\rm UV})}^2 ,
\label{eq:mass_diff_2}
\end{eqnarray}
where $\Lambda_\chi $ was introduced in such a way that,
in the high energy region $Q^2 >\Lambda_\chi^2$, 
the behavior of 
$\left[\Pi_{VV}(Q^2) - \Pi_{AA}(Q^2) \right]$ 
is well approximated by the one obtained from 
the operator product expansion (OPE) technique in the high energy region (see, e.g.,
Ref.~\cite{SVZ}).
In the real-life QCD such a scale is given by $\Lambda_\chi =4 \pi f_\pi \sim 1.1 
{\rm GeV} (>\Lambda_{\rm QCD})$, above which ($Q^2 >\Lambda_\chi^2$)
the QCD gauge coupling $\alpha$ becomes small enough for
the OPE to be obviously valid. On the other hand, in the large $N_f$ QCD 
the QCD gauge coupling $\alpha$ is almost constant (walking) over the range
$\Lambda_\chi^2 < Q^2 <\Lambda^2$ 
 and  
is not very small (of order $1$), where $\Lambda_\chi= 4\pi f_\pi \simeq 4.7\, m
\, (\ll \Lambda)$
in the large $N_f$ QCD~\cite{Harada:2003dc} ($m$ is the dynamical mass of the fermion). 
Nevertheless we expect that the OPE is still operative as was discussed~\cite{Cohen:1988sq} 
in the walking/conformal (scale-invariant) technicolor. Note that OPE obviously breaks down for $Q^2<m^2$ since 
the gauge coupling blows up due to the fact that the fermion 
which acquired mass $m$ gets decoupled from the beta function, while for $Q^2>\Lambda^2$ 
OPE is trivially valid (with vanishing anomalous dimension) 
since the gauge coupling is negligibly small due to the strong 
asymptotic freedom (see Fig.~\ref{fig:step}). 

Then the OPE dictates that the correlator $\Pi_{VV}(Q^2) - \Pi_{AA}(Q^2)$
behaves as for large $Q^2$: 
\begin{equation}
\Pi_{V-A} (Q^2) \equiv \Pi_{VV}(Q^2) - \Pi_{AA}(Q^2) \sim 
  \frac{ \langle {\bar q} q \rangle_{\Lambda_\chi}^2 }{ Q^{4-2\gamma_m}}
 \, \ \mbox{for} \ \, Q^2 > \Lambda_\chi^2 \sim (4\pi f_\pi)^2
\ ,
\label{OPE}
\end{equation}
where $\gamma_m$ is the anomalous dimension, and 
$\langle {\bar q} q \rangle_{\Lambda_\chi}$ is 
${\bar q} q$ condensation at the scale of $\Lambda_\chi$.\footnote{
Making explicit the dimensions, Eq.~(\ref{OPE}) reads $\Pi_{V-A} (Q^2)
\sim \left(\Lambda_\chi\right)^2 \cdot\left(\Lambda_\chi/Q\right)^{4-2 \gamma_m}$.
} Note that 
the large $N_f$ QCD is a walking theory characterized by 
$\gamma_m \simeq 1$~\cite{Yamawaki:1985zg}
which was confirmed by explicit calculation~\cite{Harada:2003dc}.
On the other hand, in the real-life QCD we have $\gamma_m \simeq 0$
up to logarithm.~\footnote{
In the real-life QCD the OPE in Eq.~(\ref{OPE}) actually
has a logarithmic factor $\alpha(Q^2) \times (\ln (Q^2/\Lambda_{\rm QCD}^2))^A$,
with $A=24/(33-2N_f)=8/9 \, (N_f=3)$, which however reads 
$(\ln (Q^2/\Lambda_{\rm QCD}^2))^{-1/9}$ and yields only a negligible $Q^2$ dependence.
}
Eq.~(\ref{OPE}) roughly reflects the slowly damping behavior of 
$(\Pi_{VV}(Q^2) - \Pi_{AA}(Q^2))/f_\pi^2$ of the large $N_f$ QCD compared with the real-life QCD
in Fig.~\ref{fig:largeNf_vs_realQCD.eps} at least in 
the high energy region $Q^2 > \Lambda_\chi^2 $. 
In fact, $1/Q^2$ damping for $\gamma_m\simeq 1$ yields
the integral for $\Delta m_{\pi ({\rm UV})}^2$ in Eq.~(\ref{eq:mass_diff_2}) 
a logarithmic divergence which is cutoff by $\Lambda^2$ beyond which the integrand
changes into the $1/Q^4$ damping due to  the strongly asymptotic freedom there.  
Then the enhancement by the large anomalous dimension $\gamma_m\simeq 1$ reads roughly
\begin{equation}
\frac{\Delta m_{\pi ({\rm UV})}^2}{f_\pi^2} \sim \ln (\Lambda^2/\Lambda_\chi^2)
\sim \frac{2 q}{\sqrt{\frac{\alpha_*}{\alpha_{\rm cr}}-1}} -2 \ln (4 \pi p),
\label{eq:fpi_fit}
\end{equation}
where we have used Eq.~(\ref{fpilargenf}), with $p \simeq  15.1, \, q\simeq 4.9$.
%Eq.~(\ref{fpi}), the result of Ref. \cite{Harada:2003dc}.
This implies that $\Delta m_{\pi ({\rm UV})}^2/f_\pi^2$ {\it diverges as inverse
square root scaling}, when we approach
to the critical point.  
As to the infrared contributions to $\Delta m_{\pi ({\rm UV})}^2/f_\pi^2$, 
we can see from the areas divided by the vertical line of $Q^2 =(4\pi 
f_\pi)^2$ in 
Fig.~\ref{fig:largeNf_vs_realQCD.eps}
that, even evaluated with $\Lambda_{\rm Num}
(\ll \Lambda)$, $\Delta m_{\pi ({\rm UV})}^2/f_\pi^2$ dominates $\Delta m_{\pi ({\rm IR})}^2/f_\pi^2$, 70\% to 30\%, already at $\alpha_*=0.90$, in sharp contrast to the real-life 
QCD where UV contributions vs IR contributions are opposite ratio: 20\% to 80\%.
We expect that as we approach to the critical point $\alpha_* \rightarrow \alpha_{\rm cr}$, the cutoff $\Lambda^2$ of the integral region in Fig.~\ref{fig:largeNf_vs_realQCD.eps} grows rapidly even if the damping slope were unchanged, since
the figure is drawn in the unit of $f_\pi^2$ 
which actually vanishes exponentially  with respect to $\Lambda^2$.
%Moreover, we shall later discuss the slope (i.e., $\hat{S}$) itself becomes 
%flatter to zero near the critical point.
Then  $\Delta m_{\pi ({\rm UV})}^2/f_\pi^2\rightarrow \Delta m_{\pi}^2/f_\pi^2$
near the critical point. Hence we conclude an inverse square root scaling for 
$\Delta m_{\pi}^2/f_\pi^2$:
\begin{equation}
\frac{\Delta m_{\pi}^2}{f_\pi^2} \sim 
\frac{1}{\sqrt{\frac{\alpha_*}{\alpha_{\rm cr}}-1}} .
\end{equation}

This is compared  %actually consistent 
with our numerical data in Fig.~\ref{fig:massdiff_over_fpi2}, 
%$\ln \left(\Lambda^2/\Lambda_{\rm Num}^2\right)$,
%Since the contributions from 
%the walking region should
%behave like $\ln \left(\Lambda^2/\Lambda_{\rm Num}^2\right)$, it could
%give a large contributions unless  $\Lambda^2/\Lambda_{\rm Num}^2 \sim
%constant \sim O(1)$ as we approach the critical point.
where the solid line is a fitting function as 
\begin{equation}
I(\alpha_\ast)  =  m \times
            \left(
          \frac{ 2\pi }{ \sqrt{\frac{\alpha_\ast}{\alpha_{\rm cr}} - 1} }  +  n
            \right)
            \label{inversesqrtfit}
\end{equation}
which was obtained by fitting it to the calculated data. Best fitted values for $m$ and $n$ are found 
to be $m \simeq 0.035$ and $n \simeq -0.82$.
Here we note that  Fig.~\ref{fig:massdiff_over_fpi2} was given by the contributions from the region 
$0< Q^2 < \Lambda_{\rm Num}^2$
instead of $0< Q^2 < \Lambda^2$. We shall later discuss additional
contributions from $\Lambda_{\rm Num}^2 < Q^2 < \Lambda^2$ which
may be estimated analytically based on the walking behavior of
$\Pi_{V-A}(Q^2)\sim 1/Q^2$. 
In spite of lacking the contributions from $\Lambda_{\rm Num}^2 < Q^2 < \Lambda^2$,
our data nevertheless show an enhanced mass of pseudo NG bosons reflecting an inverse square root scaling
of the walking theory:
\begin{equation}
\frac{\Delta m_{\pi}^2}{f_\pi^2} \simeq I(\alpha_\ast)
\sim  \frac{ 0.07 \pi }{ \sqrt{\frac{\alpha_\ast}{\alpha_{\rm cr}} - 1} } 
\label{inversesquare}
\end{equation}
near the critical point $\alpha_\ast =\alpha_\ast(N_f,N_c) \rightarrow 
\alpha_{\rm cr}$. 
%This implies an arbitrarily large mass for the pseudo 
%NG bosons under the model setting satisfying Eq.~(\ref{fpi}). 

\begin{figure}
  \begin{center}
    \includegraphics[height=9cm]{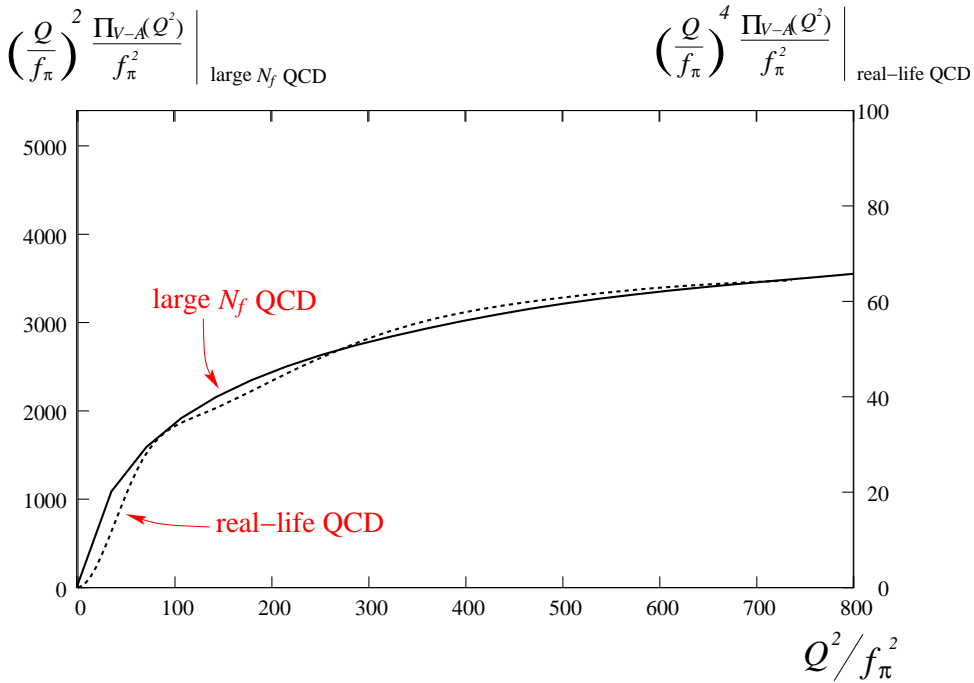}
  \end{center}
\caption{Comparison between 
$\left(\frac{Q}{f_\pi}\right)^2 \cdot \Pi_{V-A}(Q^2)/f_\pi^2$ 
in the large $N_f$ QCD with $\alpha_\ast=0.90$ (solid line) and 
$\left(\frac{Q}{f_\pi}\right)^4 \cdot \Pi_{V-A}(Q^2)/f_\pi^2$ 
in real-life QCD with $t_F = 0.04$ calculated in 
Ref.~\cite{Harada:2004qn} (dashed line). } 
\label{fig:IR_scaling}
\end{figure}

In passing, 
Eq.~(\ref{OPE}) further suggests that
\begin{eqnarray}
Q^4 \Pi_{V-A} (Q^2)\Big |_{\rm real-life QCD} 
\sim
Q^2 \Pi_{V-A} (Q^2)
\Big |_{{\rm large} N_f {\rm QCD}} 
\sim {\rm const.}
 \, \ \mbox{for} \ \, Q^2 > \Lambda_\chi^2\, . 
\end{eqnarray}
It is amusing to note that the first relation seems to hold numerically all 
the way down to the infrared region $Q^2 < \Lambda_\chi^2$
where the OPE as it stands obviously breaks down, see 
Fig.~\ref{fig:IR_scaling}.~\footnote{
The line of the large $N_f$ QCD in Fig.~\ref{fig:IR_scaling}
does not yet level off to a constant value up till this momentum region. This is 
consistent with  Ref.~\cite{Harada:2003dc} where the numerical values of 
$\gamma_m$ calculated for the values of $\alpha_*$  still away from the critical coupling
 are slightly larger than 1.
}
The remnant of the OPE carrying the information
of the anomalous dimension seems to persist as the difference of power
behaviors according to the difference of the anomalous dimension  
even in the IR region.

%%%%%%%%%%%%%%%%%%%%%%%%%%%%%%%%%%%%%%%%%%%%%%%%%%%%%%%%%%%%%%%%%%%%%
\subsection{Calculation of $S$ parameter}
\label{sec:S_result}

\begin{figure}
  \begin{center}
    \includegraphics[height=8.5cm]{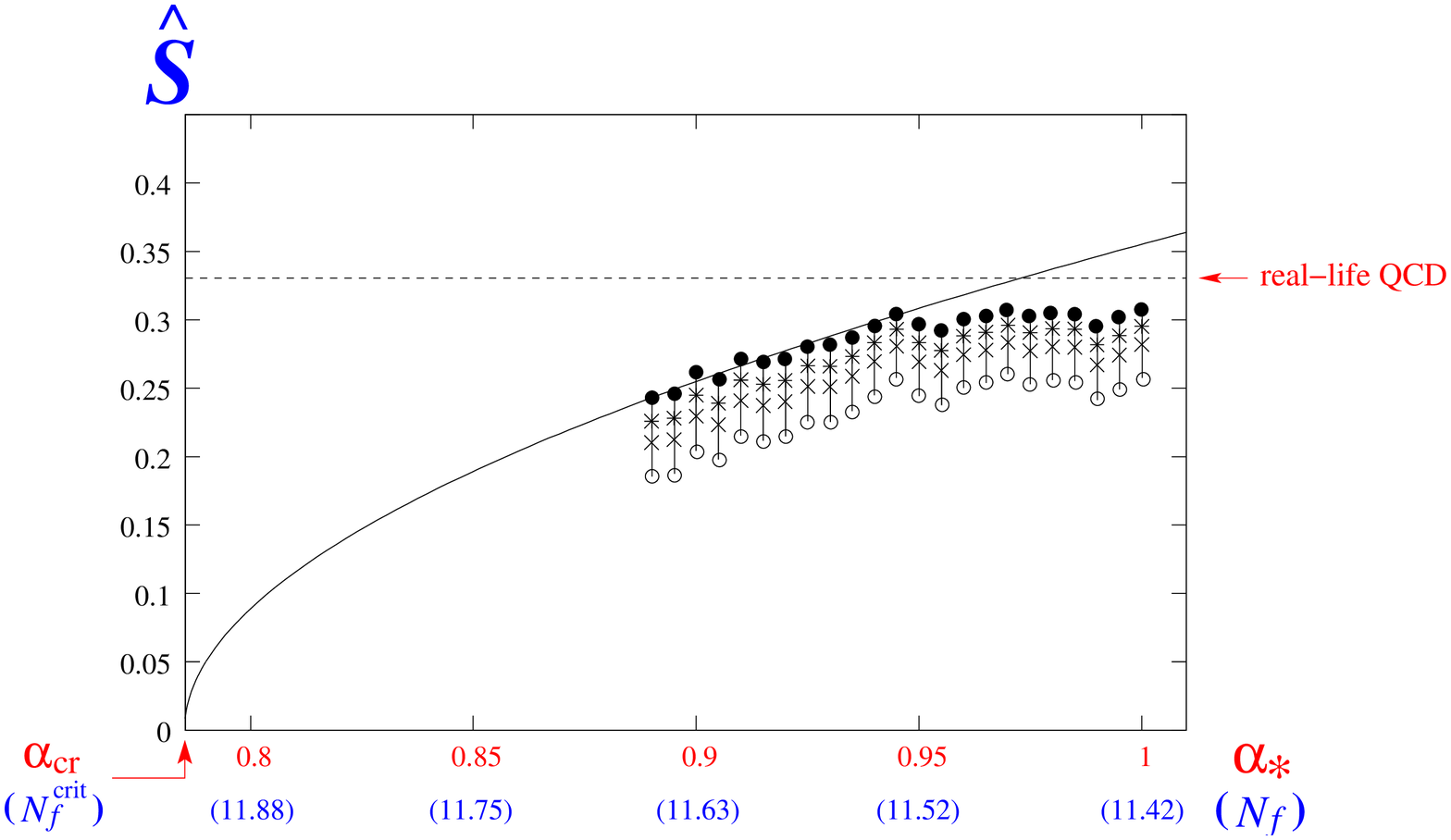}
  \end{center}
\caption{Values of ${\hat S}$ calculated from
 Eq.~(\ref{eq:S_parameter}) for  several values of $\alpha_\ast$ in the
 range of $\alpha_\ast \in [0.89:1]$.
 Points indicated by $\circ$, $\times$, $\ast$, and $\bullet$ are 
 values of ${\hat S}$ calculated by using 3, 4, 5, and 7 data points 
 of $\Pi_{VV}-\Pi_{AA}$, 
 respectively. 
Solid line shows the form of function $J(\alpha_\ast)$
in Eq.~(\ref{sqrtscaling}).
 Value of ${\hat S}$ of the real-life QCD calculated in 
 Ref.~\cite{Harada:2004qn} is also shown in the figure.
Here, we used the data 
with the infrared cutoff parameter of the running coupling 
$t_F=0.04$.} 
\label{fig:S_in_largeNf}
\end{figure}

In Fig.~\ref{fig:S_in_largeNf}, we plot ${\hat S}$ for
several values of $\alpha_\ast$ in the range of 
$\alpha_\ast \in [0.89:1]$.
There is an ambiguity how many data points we use when we calculate the 
differential coefficient of $\Pi_{VV}(Q^2) - \Pi_{AA}(Q^2)$ at $Q^2=0$.
We calculate it by using 3, 4, 5, and 7 data points, and resultant 
values of ${\hat S}$ are indicated in Fig.~\ref{fig:S_in_largeNf} 
by $\circ$, $\times$, $\ast$, and 
$\bullet$, respectively. The value of ${\hat S}$ is 
sensitive to the choice of number of data points we use, however, 
it converges as we increase the number of data points.
From this figure, we see that ${\hat S}$ in the large $N_f$
QCD near the phase transition point takes values 
\begin{equation}
 {\hat S}\ 
=0.25\quad {\rm to}\quad  0.30,
\end{equation}
which are slightly smaller (but without much difference)  
than that of the real-life QCD calculated in the same method,
${\hat S} \simeq 0.33$,  and have a tendency decreasing as we
approach the critical point.

Although our results are still larger than the estimation from the perturbative 
calculation~\cite{PeskinTakeuchi}, 
${\hat S}_{\rm pert} = \frac{N_c}{6\pi} \simeq 0.16$, 
at least in the range of $\alpha_\ast \in [0.89:1]$ investigated here,
a simple-minded extrapolation seems to suggest that it might get to 
smaller values  in the range
 of 
 $\alpha_\ast \in [\alpha_{\rm cr}:0.89]$.
 Although in the present analysis we were not able to solve the IBS equation in the range
 of 
 $\alpha_\ast \in [\alpha_{\rm cr}:0.89]$ because the numerical
 calculation near the critical point is quite difficult, 
it will motivate future work  to extend the analysis to the region closer to the
critical point and see whether or not $\hat{S}$ gets dramatic reduction near the critical 
point.   

\begin{figure}
 \begin{center}
  \includegraphics[height=8.5cm]{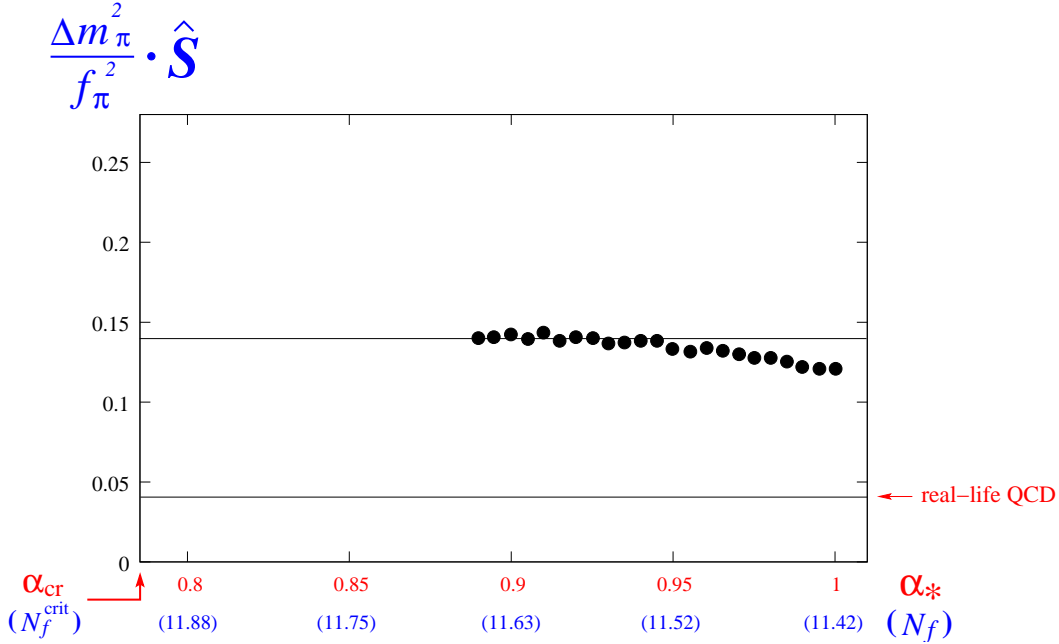}
 \end{center}
 \caption{
Values of $\hat{S}\cdot\Delta m_\pi^2/f_\pi^2$ for several values of $\alpha_\ast$.
Solid line shows the value obtained by averaging the values of data 
below $\alpha_\ast = 0.95$ as shown in Eq.~(\ref{average}).
Value of $\hat{S}\cdot\Delta m_\pi^2/f_\pi^2$ of the real-life QCD calculated in 
 Ref.~\cite{Harada:2004qn} is also shown in the figure.
Here, we used the data 
with the infrared cutoff parameter of the running coupling 
$t_F=0.04$.
 }
 \label{fig:S_times_massdiff_over_fpi2}
\end{figure}

In passing,
here we note an interesting correlation between $\hat{S}$ and the numerical
value of $\Delta m_\pi^2$ with an artificial cutoff $\Lambda_{\rm Num}$. 
In Fig.~\ref{fig:largeNf_vs_realQCD.eps}
$\Delta m_{\pi}^2/f_\pi^2$ is proportional to the area surrounded by
the vertical axis, horizontal axis, the curve of 
$(\Pi_{VV}(Q^2) - \Pi_{AA}(Q^2))/f_\pi^2$ and the line of the cutoff  $Q^2=\Lambda_{\rm Num}^2$
which does depend on the slope of the current correlators $\Pi_{V-A}(Q^2)$ at $Q^2=0$,
namely the $\hat{S}$. When  $\hat{S}$ gets smaller, $\Lambda_{\rm Num}^2$ defined by
Eq.~(\ref{eq:numerical_cutoff}) necessarily 
increases and 
so does $\Delta m_{\pi}^2/f_\pi^2$ in our numerical calculations.\footnote{
This is contrasted to the case where our integration is cut off at $\Lambda^2$ 
(instead of $\Lambda_{\rm Num}^2$) which has 
no correlation with the slope at $Q^2=0$. 
}
In Fig.~\ref{fig:S_times_massdiff_over_fpi2} we plotted 
$S\cdot\Delta m_\pi^2/f_\pi^2$ for several values of 
$\alpha_\ast$ in the range of $\alpha_\ast \in [0.89:1]$.
This product is almost constant near the critical point
and seems to stay as a constant also in the region
$\alpha_\ast \in [\alpha_{\rm cr}:0.89]$:
\begin{eqnarray}
\frac{\Delta m_\pi^2}{f_\pi^2}
\times \hat{S} 
\sim {\rm constant} \simeq 0.14 \quad (\alpha_\ast \leq 0.95)
\label{average}
\end{eqnarray}
 From this we might infer the behavior of $\hat{S}$ as an inverse function of 
$\Delta m_\pi^2 / f_\pi^2$ given in Eq.~(\ref{inversesqrtfit}):
\begin{equation}
\hat{S} \simeq J(\alpha_\ast)  = \frac{0.14}{I(\alpha_\ast)} 
=   4 \left(
             \frac{ 2\pi }{ \sqrt{\frac{\alpha_\ast}{\alpha_{\rm cr}} - 1} }  
             -0.82
           \right)^{-1}.
           \label{sqrtscaling}
\end{equation}

However, this argument is totally based on the assumption that 
$\Lambda_{\rm Num}^2/\Lambda_\chi^2$ scales in the same way 
as  $\Lambda^2/\Lambda_\chi^2$ does
in a Miransky scaling. 
Although there is no
solid theoretical support for such a continuation 
of the Miransky scaling
in the more vicinity of the critical point as 
we discuss later, 
$\Lambda_{\rm Num}^2/\Lambda_\chi^2$ imitates 
the Miransky scaling as far as the region we
explicitly calculated is concerned.
 Thus, it is certainly interesting to check in future 
 the above square root scaling of $\hat{S}$ by directly computing 
$\hat{S}$  in the region closer to the critical point.

In Ref.~\cite{Appelquist:1998xf}, it was argued 
that there arises a negative
contribution 
proportional to the dynamical mass of the fermion 
due to the
walking behavior of the running coupling, and
that the $S$ parameter in the large $N_f$ QCD could be much reduced
relative to that in QCD-like theories. 
However, as far as we analyzed in the present paper $S$ does not seem to take a 
negative value. 
%Although their claim of the dramatic reduction of the value of 
%$S$ is somewhat similar to our result, the reason for the reduction
%seems to be different.  
%In contrast to Ref.\cite{Appelquist:1998xf}, our conclusion is based on the straightforward computation without much speculations.

%%%%%%%%%%%%%%%%%%%%%%%%%%%%%%%%%%%%%%%%%%%%%%%%%%%%%%%%%%%%%%%%%%%%%%

\section{Conclusion and Discussions}
\label{sec:Conclusions}

In this paper, in the framework of the SD and the IBS equations,
we calculated $f_\pi$, ${\hat S}$, and $\Delta m_{\pi}^2$ 
on the same footing in the large $N_f$ QCD, 
through the difference between the vector current correlator
$\Pi_{VV}$ and the axial-vector current correlator $\Pi_{AA}$.

When the chiral phase transition point is approached
from the broken phase, $\alpha_\ast\rightarrow \alpha_{\rm cr}$,
$f_\pi$ and $\Delta m_\pi^2$ go to zero both with the 
essential-singularity scaling (Miransky scaling), whereas the ratio 
$\Delta m_\pi^2/f_\pi^2$ increases and actually
scales as an inverse square root (Fig.~\ref{fig:massdiff_over_fpi2} and Eq.~(\ref{inversesquare})):
\begin{eqnarray}
 %f_\pi &\sim& 1.5 \,\Lambda \,\exp 
  %     \left(
  %          - \frac{\pi}{\sqrt{\frac{\alpha\ast}{\alpha_{\rm cr}}-1\ }}
  %     \right)
  % \ , 
  % \nonumber \\
 \frac{\Delta m_{\pi}^2}{f_\pi^2} 
&\sim&
  \frac{ 0.07 \pi }{ \sqrt{\frac{\alpha_\ast}{\alpha_{\rm cr}} - 1} }   \, .
\end{eqnarray}
Thus the radiative mass of the pseudo NG boson can be dramatically enhanced
compared with the conventional estimate of the techni pseudo's based on a 
simple scale-up of QCD with a  factor scaling of $N_f$ and $N_c$.

Here we should mention that our numerical estimate of  $\Delta m_\pi^2/f_\pi^2$ 
was done on the integral for the region $0<\, Q^2\, <\,\Lambda_{\rm Num}^2$  instead
of $0<\, Q^2\, <\,\Lambda^2$, since it is rather difficult to 
estimate them by directly solving the IBS equation numerically including larger $Q^2$
regions. We now
discuss the additional contributions from
$\Lambda_{\rm Num}^2<\, Q^2\, <\,\Lambda^2$ through an
analytical estimate, instead of directly solving the IBS equation, 
based on the observed walking behavior of the current correlators
$\Pi_{V-A}(Q^2) \sim 1/Q^2$. We divide the integral for $\Delta m_{\pi}^2$ as 
\begin{eqnarray}
   \Delta m_{\pi}^2 
   &=& \frac{3 \alpha_{em}}{4 \pi f_\pi^2} 
     \left(
       \int_{0}^{\Lambda_{\rm Num}^2} d Q^2 +
       \int_{\Lambda_{\rm Num}^2}^{\Lambda^2} d Q^2 
     \right)
   \left[ \Pi_{V-A}(Q^2) \right] ,\\
&\equiv&    \Delta m_{\pi({\rm Num})}^2 +   \Delta m_{\pi({\rm UV >})}^2.
\end{eqnarray}
By definition of $\Lambda_{\rm Num}$ in 
Eq.~(\ref{eq:numerical_cutoff})
we have
\begin{equation}
\frac{\Pi_{V-A}(Q^2)}{f_\pi^2}  = \frac{1}{50} \frac{\Lambda_{\rm Num}^2}{Q^2}
\quad  (\Lambda_{\rm Num}^2<\, Q^2\, <\,\Lambda^2) \, ,
\end{equation}
which is substituted into the above  $\Delta m_{\pi({\rm UV >})}^2$:
\begin{equation}
\frac{\Delta m_{\pi({\rm UV>})}^2}{f_\pi^2} = \frac{1}{50}\ 
  \frac{3 \alpha_{em}}{4 \pi}\ 
  \frac{\Lambda_{\rm Num}^2}{f_\pi^2} \ 
   \ln{\left(
     \frac{\Lambda^2}{\Lambda_{\rm Num}^2}
       \right)}.
\end{equation}

\begin{table}[h]
\begin{center}
 \begin{tabular}{|c||c|c||c||c|c|}  \hline
$\alpha_\ast$ & $\Lambda_{\rm Num}^2/\Lambda_\chi^2$
 &$\Lambda^2/\Lambda_{\rm Num}^2$ &
$\alpha_\ast$ & $\Lambda_{\rm Num}^2/\Lambda_\chi^2$
 &$\Lambda^2/\Lambda_{\rm Num}^2$ \\ 
\hline\hline 

   0.890 & $ 28 $  & $ 3.8 \times 10^{5}$ &   0.950 & $ 21 $  & $ 2.7 \times 10^{3}$\\ \hline 
   0.895 & $ 29 $  & $ 2.0 \times 10^{5}$ &   0.955 & $ 21 $  & $ 1.9 \times 10^{3}$\\ \hline 
   0.900 & $ 26 $  & $ 1.3 \times 10^{5}$ &   0.960 & $ 21 $  & $ 1.4 \times 10^{3}$\\ \hline 
   0.905 & $ 27 $  & $ 7.6 \times 10^{4}$ &   0.965 & $ 21 $  & $ 1.1 \times 10^{3}$\\ \hline 
   0.910 & $ 26 $  & $ 4.8 \times 10^{4}$ &   0.970 & $ 20 $  & $ 8.8 \times 10^{2}$\\ \hline 
   0.915 & $ 25 $  & $ 3.3 \times 10^{4}$ &   0.975 & $ 20 $  & $ 6.7 \times 10^{2}$\\ \hline 
   0.920 & $ 26 $  & $ 2.0 \times 10^{4}$ &   0.980 & $ 20 $  & $ 5.2 \times 10^{2}$\\ \hline 
   0.925 & $ 25 $  & $ 1.4 \times 10^{4}$ &   0.985 & $ 19 $  & $ 4.1 \times 10^{2}$\\ \hline 
   0.930 & $ 23 $  & $ 1.0 \times 10^{4}$ &   0.990 & $ 19 $  & $ 3.3 \times 10^{2}$\\ \hline 
   0.935 & $ 23 $  & $ 7.0 \times 10^{3}$ &   0.995 & $ 18 $  & $ 2.7 \times 10^{2}$\\ \hline 
   0.940 & $ 23 $  & $ 4.9 \times 10^{3}$ &   1.000 & $ 18 $  & $ 2.2 \times 10^{2}$\\ \hline 
   0.945 & $ 22 $  & $ 3.6 \times 10^{3}$ &&&\\ \hline 

 \end{tabular}
\end{center}
\caption{Values of $(\Lambda_{\rm Num}^2/\Lambda_\chi^2)$ and 
$(\Lambda^2/\Lambda_{\rm Num}^2)$
for several values of $\alpha_\ast$.}
\label{tab:1}
\end{table}

Values of $(\Lambda_{\rm Num}^2/\Lambda_\chi^2)$ and  
$(\Lambda^2/\Lambda_{\rm Num}^2)$ are summarized in  Table~\ref{tab:1}.
Considering the definition of $\Lambda_\chi$ ($\equiv 4\pi f_\pi$) 
and the fact that the scaling of $f_\pi/\Lambda$ is identified as 
Miransky scaling (see Eq.~(\ref{eq:fpi_fit}) and Fig.~\ref{fig:fpi_in_largeNf}), 
we can understand that $\Lambda^2/\Lambda_\chi^2$ also shows 
the (inverse) Miransky scaling.
$\Lambda^2/\Lambda_\chi^2$ is viewed as a 
product of $\Lambda^2/\Lambda_{\rm Num}^2$ and 
$\Lambda_{\rm Num}^2/\Lambda_\chi^2$, the former becoming more prominent
as we get closer to the critical point. This implies that cutting off the integral
for $\Delta m_\pi^2$ at $\Lambda_{\rm Num}^2$ does not reproduce main part of the walking 
theory and hence does not yield a good approximation,
in sharp contrast to the case of the real-life QCD.
Nevertheless $\Lambda_{\rm Num}^2/\Lambda_\chi^2$ is also
growing  as we 
get closer to the critical point, which corresponds to  decreasing of $\hat{S}$.
This  actually 
imitates the Miransky scaling in such a way that $\Delta m_{\pi({\rm Num})}^2$
behaves as if in an inverse square root scaling observed in Fig.~\ref{fig:massdiff_over_fpi2} and 
Eq.~(\ref{inversesquare}). 

\begin{figure}[h]
  \begin{center}
    \includegraphics[height=9cm]{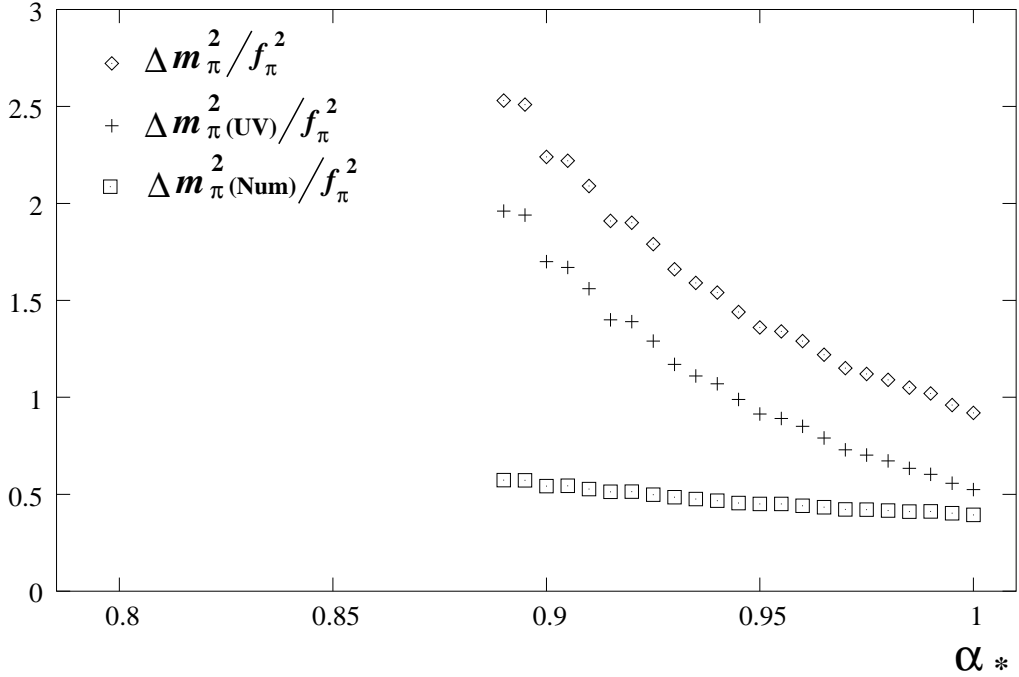}
  \end{center}
\caption{Values of $\Delta m_{\pi({\rm Num})}^2$, 
$\Delta m_{\pi({\rm UV>})}^2$, and $\Delta m_{\pi}^2$ in the unit of
$f_\pi^2$ for several 
values of $\alpha_\ast$.}
\label{fig:delta_mpi2_over_fpi2_new}
\end{figure}

In Fig.~\ref{fig:delta_mpi2_over_fpi2_new}, 
we plot the values of $\Delta m_{\pi({\rm Num})}^2$, 
$\Delta m_{\pi({\rm UV>})}^2$, and $\Delta m_{\pi}^2$ in the unit of
$f_\pi^2$ for several 
values of $\alpha_\ast$. As was noted above, the additional contributions from $\Lambda_{\rm Num}^2
< Q^2 < \Lambda^2$ is much larger than those from $Q^2 <\Lambda_{\rm Num}^2$ which we 
calculated numerically, since  $\Lambda^2/\Lambda_{\rm Num}^2  \gg \Lambda_{\rm Num}^2/\Lambda_\chi^2$
(see Table \ref{tab:1}) in the walking regions.
The fitting function for total $\Delta m_{\pi}^2$ reads 
$m\simeq 0.042$ and $n\simeq -0.268$ in the parameterization of
 Eq.~(\ref{inversesqrtfit}).
This figure clearly shows that the contribution 
from $Q^2>\Lambda_{\rm Num}^2$ dominates the scaling of total $\Delta m_{\pi}^2$.
When it is applied for a popular technicolor model (one-family model) of Farhi-Susskind~\cite{techni},
with $\Lambda_\chi = 4 \pi f_\pi\sim 4\pi \times  123 {\rm GeV}$, we are interested in the
region $\Lambda< \Lambda_{\rm ETC}\sim 10^3 {\rm TeV}$ and hence $\Lambda/\Lambda_\chi
\sim 6.5\times 10^2$ (we are using $N_{\rm TC}=3$ in this case). This roughly corresponds to 
$\alpha_* =0.975$ in Table \ref{tab:1} and $\Delta m_\pi^2 /f_\pi^2 =1.12$, about 9 times larger than the value 
estimated in
 the real-life QCD
$\Delta m_\pi^2 /f_\pi^2\simeq 0.123$~\cite{Harada:2004qn} (experimental value $\Delta m_\pi^2
=0.148\pm 0.001$)  in the same method. This suggests that the radiative mass of the pseudo
NG bosons in the walking technicolor  could be enhanced by a factor 3
compared with a simple scale up of the QCD. The typical estimate of $100 {\rm GeV}$ mass range~\cite{techni}
would be boosted to $300 {\rm GeV}$ range, which could be of phenomenological relevance.

%On the other hand,
As to $\hat{S}$, our numerical result showed
${\hat S}\simeq 0.25-0.30$ which is somewhat smaller than that in the real-life QCD and 
indicated a decreasing tendency  
as we approach the critical point. 
Here we note that there is a subtlety to evaluate the slope of $\Pi_{V-A}(Q^2)$ at $Q^2=0$
numerically. To check this subtlety we compute another quantity $\hat{S}_*$:
\begin{equation}
{\hat S}_* \equiv -4\pi\ \frac{\Pi_{V-A}(Q^2=f_\pi^2) 
- \Pi_{V-A}(Q^2=0)}{f_\pi^2},
\end{equation}
which is similar to the popular definition of $S$ except for the point that ours is defined
for the space-like momentum instead of the time-like one.
Numerical results for several values of $\alpha_\ast$ 
are summarized in Table~\ref{tab:2}. 
\begin{table}[h]
\begin{center}
 \begin{tabular}{|c||c|c|c|}  \hline
$\alpha_\ast$ & $\Pi_{V-A}(Q^2=0)/\Lambda^2=f_\pi^2/\Lambda^2$
 &$\Pi_{V-A}(Q^2=f_\pi^2)/\Lambda^2$ & ${\hat S}_*$ \\ \hline\hline 

   0.890 & $6.05 \times 10^{-10}$ & $5.90 \times 10^{-10}$ & 0.308  \\ \hline 

   0.895 & $1.08 \times 10^{-9}$  & $1.05 \times 10^{-9}$  & 0.309  \\ \hline 

   0.900 & $1.85 \times 10^{-9}$  & $1.80 \times 10^{-9}$  & 0.311  \\ \hline 

   0.905 & $3.09 \times 10^{-9}$  & $3.01 \times 10^{-9}$  & 0.311  \\ \hline 

   0.910 & $5.00 \times 10^{-9}$  & $4.88 \times 10^{-9}$  & 0.311  \\ \hline 

   0.915 & $7.91 \times 10^{-9}$  & $7.71 \times 10^{-9}$  & 0.312  \\ \hline 
 
   0.920 & $1.22 \times 10^{-8}$  & $1.19 \times 10^{-8}$  & 0.313  \\ \hline 

   0.925 & $1.85 \times 10^{-8}$  & $1.80 \times 10^{-8}$  & 0.314  \\ \hline 

   0.930 & $2.74 \times 10^{-8}$  & $2.67 \times 10^{-8}$  & 0.314  \\ \hline 

   0.935 & $3.99 \times 10^{-8}$  & $3.89 \times 10^{-8}$  & 0.315  \\ \hline 

   0.940 & $5.72 \times 10^{-8}$  & $5.57 \times 10^{-8}$  & 0.316  \\ \hline 

   0.945 & $8.06 \times 10^{-8}$  & $7.86 \times 10^{-8}$  & 0.316  \\ \hline 

   0.950 & $1.12 \times 10^{-7}$  & $1.09 \times 10^{-7}$  & 0.317  \\ \hline 

   0.955 & $1.54 \times 10^{-7}$  & $1.50 \times 10^{-7}$  & 0.318  \\ \hline 

   0.960 & $2.08 \times 10^{-7}$  & $2.03 \times 10^{-7}$  & 0.318  \\ \hline 

   0.965 & $2.78 \times 10^{-7}$  & $2.71 \times 10^{-7}$  & 0.319  \\ \hline 

   0.970 & $3.68 \times 10^{-7}$  & $3.59 \times 10^{-7}$  & 0.320  \\ \hline 

   0.975 & $4.82 \times 10^{-7}$  & $4.70 \times 10^{-7}$  & 0.320  \\ \hline 

   0.980 & $6.25 \times 10^{-7}$  & $6.09 \times 10^{-7}$  & 0.321  \\ \hline 

   0.985 & $8.02 \times 10^{-7}$  & $7.82 \times 10^{-7}$  & 0.321  \\ \hline 

   0.990 & $1.02 \times 10^{-6}$  & $9.96 \times 10^{-7}$  & 0.322  \\ \hline 

   0.995 & $1.29 \times 10^{-6}$  & $1.26 \times 10^{-6}$  & 0.322  \\ \hline 

   1.000 & $1.62 \times 10^{-6}$  & $1.57 \times 10^{-6}$  & 0.323  \\ \hline 

 \end{tabular}
\end{center}
\caption{Values of 
$\Pi_{V-A}(Q^2=0)/\Lambda^2=f_\pi^2/\Lambda^2$,  
$\Pi_{V-A}(Q^2=f_\pi^2)/\Lambda^2$, and ${\hat S}_*$ 
for several values of $\alpha_\ast$.}
\label{tab:2}
\end{table}

\begin{figure}[h]
  \begin{center}
    \includegraphics[height=8cm]{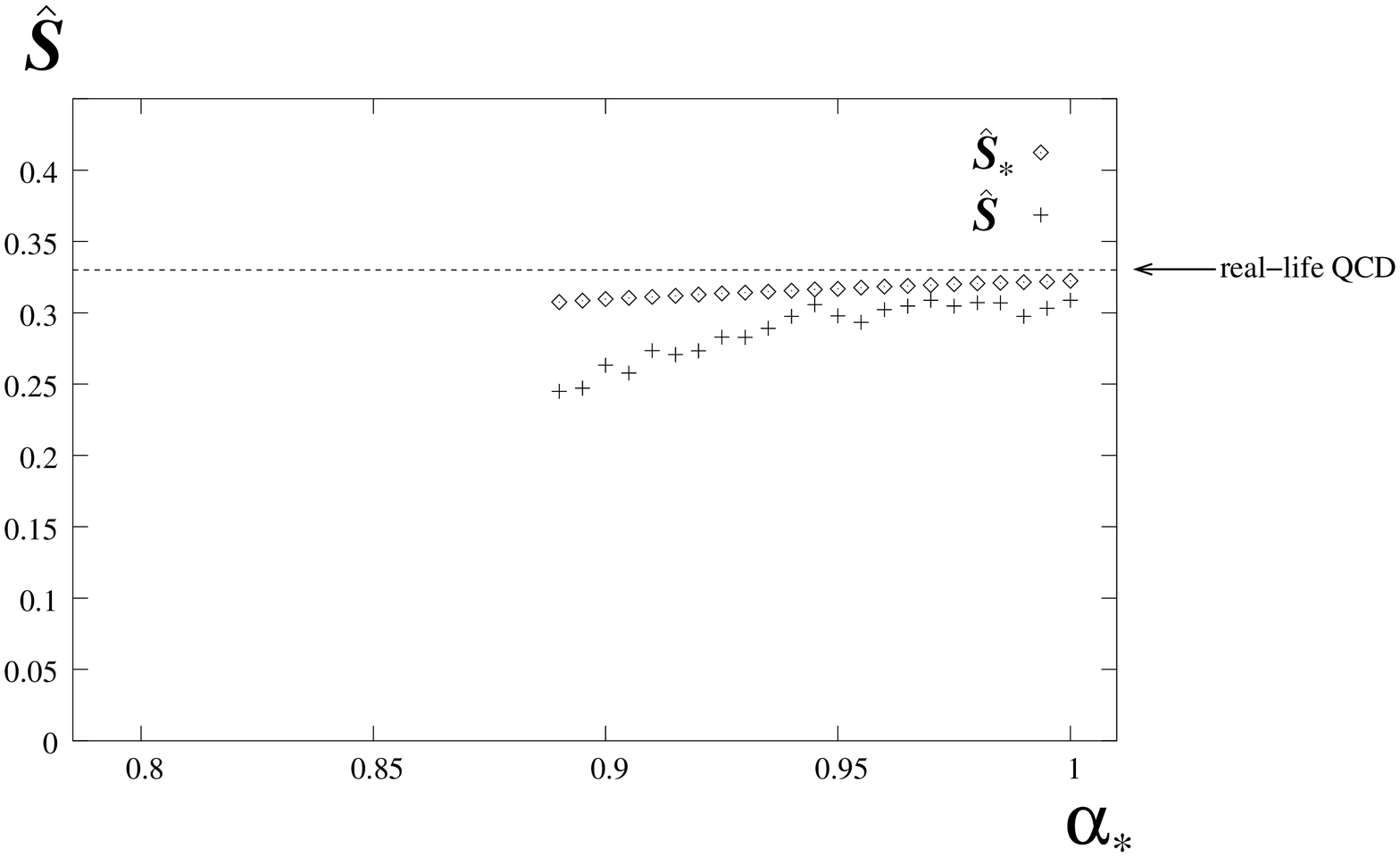}
  \end{center}
\caption{Values of ${\hat S}_*$ shown in 
Table~\ref{tab:2}. Values of ${\hat S}$ are also plotted in the figure.
}
\label{fig:S_new}
\end{figure}

In Fig.~\ref{fig:S_new}, we plotted the values of ${\hat S}_*$ shown in 
Table~\ref{tab:2}. 
The data in the previous definition are also plotted in this figure.
The result is somewhat larger than the value in the previous method
and  seems to indicate slower decreasing behavior than the previous one 
as we approach the critical
point. Thus, up to this subtlety, it seems  unlikely that $\hat{S}$ is dramatically
reduced in the large $N_f$ QCD at least in the region we calculated.
%with the product 
%$\hat{S}\cdot \Delta m_\pi^2/f_\pi^2$ being kept constant:
%\begin{eqnarray}
%\frac{\Delta m_\pi^2}{f_\pi^2}
%\times \hat{S} 
%\sim {\rm constant} \simeq 0.14 \quad (\alpha_\ast \leq 0.95).
%\end{eqnarray}
%This implies that $S$ goes to zero in a square root scaling:
%\begin{eqnarray}
%S =\frac{N_c}{3}\frac{N_f}{2} \hat{S}\sim \frac{N_c}{3}\frac{N_f}{\pi} \sqrt{\frac{\alpha_\ast}{\alpha_{\rm cr}} - 1} \rightarrow \, 0.
%\end{eqnarray}
%Thus we found a simple solution to the long-standing $S$ parameter
%problem of the technicolor. By arranging $N_f$ and $N_c$ in such a way that
%the theory is close to the critical point, $\alpha_\ast \rightarrow \alpha_{\rm cr}$, we can easily get $S$ parameter to satisfy the precision
%electroweak measurements ($S \lesssim 0.1$ \cite{Eidelman:2004wy}).
At any rate it is highly desirable to extend our analysis to
closer region to the critical point and check by explicit computation
whether or not our conclusion in this paper  persists there. 

In the present analysis, we only considered $SU(N_c)$ gauge theory with 
$N_c=3$ (with $N_f$ flavors). This is because, by taking $N_c=3$, it
became  
possible to compare the result of the large 
$N_f$ $SU(N_c)$ gauge theory with usual QCD ($N_c=3, N_f=3$), and see the 
difference of dynamics which is caused by the difference of the number
of flavors. 
However, there is no reason to prohibit us from considering the large
$N_f$ $SU(N_c)$ gauge theory with $N_c \neq 3$ as an underlying theory of
the electroweak symmetry breaking.
(As an example of recent work on walking technicolor model 
based on $SU(2)$ gauge theory, see \cite{Dietrich:2005jn,Christensen:2005cb}.)
So it is interesting to investigate ${\hat S}$ and $\Delta m_{\pi}^2$ 
in the case of $N_c\neq 3$ in future works.

%In conclusion, based on our analysis, the long standing problem with the 
%$S$ parameter is no longer fatal to 
%the technicolor model building. 

%%%%%%%%%%%%%%%%%%%%%%%%%%%%%%%
%%%%%%%%%%%%%%%%%%%%%%%%%%%%%%%%%%%%%%%%%%%%%%%%%%%%%%%%%%%%%%%%%%%%

\section*{Acknowledgments}
We would like to thank Robert Shrock for very illuminating discussions and
criticism. We also thank Masaharu Tanabashi for useful discussions. 
This work was supported in part by 
the JSPS Grant-in-Aid for the Scientific Research 
(B)(2) 14340072, 
The Mitsubishi Foundation
and by the 21st Century COE Program 
of Nagoya University provided by JSPS (15COEG01). This work was also 
supported in pat by (C)(2) 16540241 (M.H.), the Daiko Foundation \#9099 (M.H.), 
the MEXT Grant-in-Aid for Scientific Research No.14046201 (M.K.) 
and the grant NSF-PHY-00-98527 at YITP, SUNY Stony Brook (M.K.).
%M.K. would like to thank Masaharu Tanabashi and Robert Shrock for valuable discussions, and 
M.K. acknowledges the hospitality of the Aspen Center for Physics 
where part of this work was done.

%%%%%%%%%%%%%%%%%%%%%%%%%%%%%%%%%%%%%%%%%%%%%%%%%%%%%%%%%%%%%%%%%%%%
%%%%%%%%%%%%%%%%%%%%%%%%%%%%%%%%%%%%%%%%%%%%%%%%%%%%%%%%%%%%%%%%%%%%
\appendix

\renewcommand\theequation{\Alph{section}.\arabic{equation}}

\begin{flushleft}
\LARGE\bf Appendices
\end{flushleft}

%%%%%%%%%%%%%%%%%%%%%%%%%%%%%%%%%%%%%%%%%%%%%%%%%%%%%%%%%%%%%%%%%%%%
\section{Appendix : SD equation in the improved ladder approximation}
\label{app:SD}

\begin{figure}
  \begin{center}
    \includegraphics[height=3.3cm]{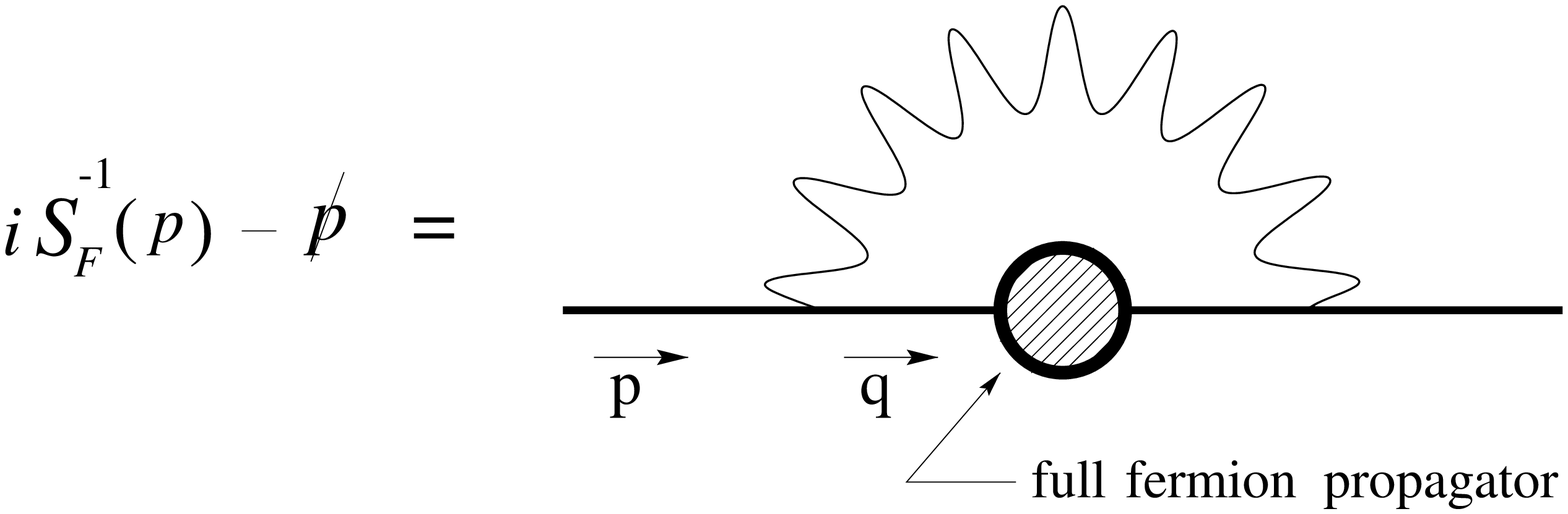}
  \end{center}
\caption{A graphical expression of the SD equation 
         in the (improved) ladder approximation.}
\label{fig:SDeq}
\end{figure}

Schwinger-Dyson (SD) equation is a powerful tool to study the 
dynamical generation of the fermion mass directly from QCD
(for reviews, see, e.g., Refs.~\cite{Kugo:1991da,Miransky:book}).
The SD equation for the full fermion propagator 
$\ i S_{F}^{-1} = A(p^2) \fsl{p} - B(p^2)\ $ in the improved 
ladder approximation~\cite{Miransky:vj} 
is given by 
(see Fig.~\ref{fig:SDeq} for a graphical expression)
\begin{equation}
  i S_F^{-1}(p) - \fsl{p}\ =\  C_2\int \frac{d^4 q}{i (2 \pi)^4}\ 
  \bar{g}^2(p,q)\ \frac{1}{(p-q)^2}
  \left( g_{\mu \nu} - \frac{(p-q)_\mu(p-q)_\nu}{(p-q)^2} \right)
  \gamma^\mu \ i S_F(q) \ \gamma^\nu ,
\label{eq:improved_ladder_SD_0}
\end{equation}
where $\ C_2 \left(  = \frac{N^2_c - 1}{2 N_c} \right) \ $
 is the second casimir invariant, $\ \bar{g}(p,q)\ $ is 
the  running coupling, and the
Landau gauge is adopted for the gauge boson propagator.
The SD equation provides coupled equations for
two functions $A$ and $B$ in 
the full fermion propagator $S_{F}$.
When we adopt a simple ansatz for the 
running coupling, $ \bar{g}^2(p,q) =\bar{g}^2(\max (p_E^2, q_E^2))
$~\cite{Miransky:vj},
with $(p_E^2, q_E^2)$ being the Euclidean momenta,
we can carry out the angular integration and get $A(p^2)\equiv 1$
in the Landau gauge. Then the SD equation becomes a self-consistent 
equation for
the mass function $\Sigma(p^2) \equiv B(p^2)$. The resultant
asymptotic behavior of the dynamical mass 
$\Sigma(p^2)$ 
is shown to coincide
with that obtained by the operator product expansion 
technique.

However, 
it was shown in Ref.~\cite{Kugo:1992pr} that the axial Ward-Takahashi 
identity is violated in the improved ladder approximation unless the
gluon momentum is used as the argument of the running coupling
as $\bar{g}^2( (p_E-q_E)^2 )$.
In this choice
we cannot
carry out the angle integration 
analytically since the running coupling depends on the 
angle factor 
$\cos{\theta} = p_E \cdot q_E / \vert p_E \vert \vert q_E \vert $. 
Furthermore, we would need to introduce a nonlocal gauge 
fixing~\cite{Georgi:1989cd,Kugo:1992pr} to preserve the condition $A=1$.

In Ref.~\cite{Aoki:1990aq}, however, it was shown that 
an angle averaged form
$\bar{g}^2(p_E^2+q_E^2)$ 
gives a good approximation.
Then, in the present analysis
we take the argument of the running coupling as
\begin{equation}
  \bar{g}^2(p_E,q_E) \ \Rightarrow\  \bar{g}^2(p_E^2+q_E^2) .
\label{eq:approx_of_argument}
\end{equation}
After applying this angle average
approximation and carrying out the angular
integration, 
we can show (see, e.g., 
Refs.~\cite{Miransky:book})
that $A$ always satisfies $A(p^2)=1$ in the Landau gauge.
Then the SD equation becomes 
\begin{equation}
  \Sigma(x) = C_2\  \frac{3}{16 \pi^2} \int dy\  
  \frac{y\  \Sigma(y)}{y + \Sigma^2(y)} 
  \ \frac{\bar{g}^2(x+y)}{\max(x,y)} \ ,
\label{eq:improved_ladder_SD}
\end{equation}
where $\ x = p_E^2 \ $ and $\ y = q_E^2 \ $.

%%%%%%%%%%%%%%%%%%%%%%%%%%%%%%%%%%%%%%%%%%%%%%%%%%%%%%%%%%%%%%%%%%%%
\section{Appendix : Numerical method for solving the SD equation}
\label{app:num_SD}

In this appendix we briefly explain how to solve the SD equation
numerically.

We first introduce the infrared
(IR) cutoff $\lambda_{SD}$ and ultraviolet (UV) cutoff 
$\Lambda_{SD}$ as
\begin{equation}
\Lambda^2 \ e^{\lambda_{SD}/\Lambda} \ \le\  x \, , \ y \ \le \ 
\Lambda^2 \ e^{\Lambda_{SD}/\Lambda}
\ .
\end{equation}
Then, we discretize the momentum variable $x$ and  
$y$ into $N_{SD}$ points as
\begin{equation}
  x_i =  \Lambda^2 \ \exp \Big[ \lambda_{SD}/\Lambda +  
                       D_{SD}\cdot i\,
                       \Big] \ , \ \ \ 
  \Big( i = 0, 1, 2, \cdots, (N_{SD}-1) \Big) \ ,
\end{equation}
where
\begin{equation}
  D_{SD} = \frac{\ ( \Lambda_{SD} - \lambda_{SD} )/\Lambda\ }{N_{SD} - 1} \ .
\end{equation}
Accordingly, the integration over $y$ is replaced with a summation as
\begin{equation}
\int d y \ \Rightarrow \ D_{SD} \sum_{j} y_j \ .
\end{equation}
Then, the SD equation in Eq.~(\ref{eq:improved_ladder_SD}) with the
running coupling in Eq.~(\ref{eq:approx_of_argument}) is rewritten as
\begin{equation}
  \Sigma(x_i) \ =  \ \frac{1}{4 \pi^2}  
         \ D_{SD}\sum_j \  
         \bar g^2(x_i + y_j) \
         \frac{y_j^2}{\max(x_i,\ y_j)}
         \ \frac{\Sigma(y_j)}
         {y_j + \Sigma^2(y_j)} .
\label{eq:discretized_largeNf_SD_1}
\end{equation}
This discretized version of the
SD equation is solved 
by the recursion relation:
\begin{equation}
  \Sigma_{(n+1)}(x_i) =  \frac{1}{4 \pi^2}  
          D_{SD}\sum_j 
          \bar g^2(x_i + y_j) \
          \frac{y_j^2}{\max(x_i,\ y_j)}
          \ \frac{\Sigma_{(n)}(y_j)}
          {y_j + \Sigma_{(n)}^2(y_j)} \ .
\label{eq:discretized_largeNf_SD_2}
\end{equation}
Starting from 
a suitable initial condition (we choose
$\Sigma_{(0)}(x_i) = 1 $),
we update the mass function by the above recursion relation.
Then, we stop the iteration when the convergence condition 
\begin{equation}
  D_{SD}\sum_i
       \frac{x^2_i}{16 \pi^2} \,
       \Big[\,\Sigma_{(n+1)}(x_i) 
               - \Sigma_{(n)}(x_i)\,\Big]^2 
  \ <\  \varepsilon^2 \Lambda^6
\end{equation}
is satisfied for sufficiently small $\varepsilon$,
and
regard this $\Sigma_{(n)}$ as a solution of 
Eq.~(\ref{eq:discretized_largeNf_SD_1}).

In the present paper, we use the following parameters to solve 
the SD equation numerically:
\begin{equation}
  \Lambda_{SD}/\Lambda = + 24 \ \ ,\ \ 
  \lambda_{SD}/\Lambda =  - 24 \ \ ,\ \ 
  N_{SD}  = 1000 \ \ ,\ \ 
  \varepsilon = 10^{-20} .
\end{equation}

%%%%%%%%%%%%%%%%%%%%%%%%%%%%%%%%%%%%%%%%%%%%%%%%%%%%%%%%%%%%%%%%%%%%
\section{Appendix : Numerical method for solving the IBS equation}
\label{app:num_IBS}
 
 In this appendix we transform the IBS equation in 
 Eq.~(\ref{eq:IBSeq}) into the form 
 in which we can solve it numerically.
 
 First, we introduce 
 the conjugate bispinor bases defined by
 \begin{equation}
   \bar\Gamma^{(J)}_i(p;q,\epsilon) \equiv
   \gamma_0 \Gamma^{(J)}_i(p^\ast;q,\epsilon)^\dag \gamma_0 \ .
 \end{equation}
 Multiplying 
 these conjugate bispinor bases 
 from left, taking the trace 
 of spinor indices and summing over the polarizations, 
 we rewrite
 Eq.~(\ref{eq:IBSeq}) into the following form:
 \begin{equation}
   T^{(J)}_{ij}(u,x) \chi^{(J)}_j(u,x) - \frac{1}{8 \pi^3} 
   \int_{-\infty}^{\infty}dv \int_0^\infty dy y^2 
   K^{(J)}_{ij}(u,x;v,y) \chi^{(J)}_j(v,y) 
   \ =\ I^{(J)}_i(u,x) ,
 \end{equation}
 where 
 the summation over the index $j$ is understood, and
 \begin{eqnarray}
   I_i^{(J)} &=& \sum_{\epsilon} 
         \mbox{tr} \left[ \bar\Gamma_i^{(J)}(p;q,\epsilon)
 	\left( \epsilon \cdot G^{(J)}\right) \right] \ ,\\
   T^{(J)}_{ij}(u,x) &=& \sum_{\epsilon} 
         \mbox{tr} \left[ 
 	\bar\Gamma_i^{(J)}(p;q,\epsilon) T(p;q) \Gamma_j^{(J)}
           (p;q,\epsilon)
 	\right] ,\\
   K^{(J)}_{ij}(u,x;v,y) &=& \int_{-1}^{1} d\cos\theta 
          \ \sum_{\epsilon} 
         \mbox{tr} \left[ \bar\Gamma_i^{(J)}(p;q,\epsilon)
 	K(p,k) \Gamma_j^{(J)}(k;q,\epsilon) \right] \ ,
 \end{eqnarray}
 with the real variables $v$ and $y$ introduced as
 \begin{equation}
   k\cdot q = - v\ Q ,\ \ \  k\cdot p= - u v - x y \cos\theta .
 \end{equation}
 Here $\theta$ is the angle between the spatial components of 
 $p_\mu$ and $k_\mu$.
 
 Using the property of $\chi_i^{(J)}$ in Eq.~(\ref{eq:even-property}), 
 we restrict the integration range as $v > 0$:
 \begin{equation}
   \int dv K_{ij}(u,x;v,y) \chi_j^{(J)}(v,y) = 
 	\int_{v > 0} dv \left[ K_{ij}(u,x;v,y) +
 	 K_{ij}(u,x;-v,y) \right] \chi_j^{(J)}(v,y).
 \end{equation}
 Then, in the following, we treat
 all the variables $u$, $x$, $v$ and $y$
 as positive values.
 
 To discretize the
 variables $u$, $x$, $v$ and $y$
 we introduce new variables 
 $U$, $X$, $V$ and $Y$ as
 \begin{eqnarray}
   u = e^{U}\ , & & x = e^{X} \ , \nonumber\\
   v = e^{V}\ , & & y = e^{Y} \ ,
 \end{eqnarray}
 and
 set ultraviolet (UV) and infrared (IR) cutoffs as
 \begin{equation}
   U, V \  \in \  [\lambda_U,\Lambda_U] ,\ \ 
   X, Y \  \in \  [\lambda_X,\Lambda_X] .
 \end{equation}
 We discretize the variables $U$ and $V$ 
 into $N_{BS,U}$ points evenly, and $X$ and $Y$
 into $N_{BS,X}$ points.
 Then,
 the original variables are labeled as
 \begin{eqnarray}
   & & u_{[I_U]} = \exp\left[\lambda_U + D_U I_U \right], \ \ 
       x_{[I_X]} = \exp\left[\lambda_X + D_X I_X \right], \nonumber\\
   & & v_{[I_V]} = \exp\left[\lambda_U + D_U I_V \right], \ \ 
       y_{[I_Y]} = \exp\left[\lambda_X + D_X I_Y \right], \nonumber
 \end{eqnarray}
 where
 $I_U, I_V = 0, 1, 2, \cdots (N_{BS,U}-1)$ and
 $I_X, I_Y = 0, 1, 2, \cdots (N_{BS,X}-1)$.
 The measures $D_U$ and $D_X$ are defined as
 \begin{equation}
   D_U = \frac{\Lambda_U - \lambda_U}{N_{BS,U} - 1}\ ,\ \ 
   D_X = \frac{\Lambda_X - \lambda_X}{N_{BS,X} - 1} \ .
 \end{equation}
 As a result, the integration is converted into the 
 summation:
 \begin{equation}
   \int_{v > 0} y^2 \ dy\ dv\ \cdots  \ \ \  \Longrightarrow \ \ \  
   D_U D_V \sum_{I_V,I_Y} v y^3 \ \cdots.
 \end{equation}
 In
 order to avoid integrable singularities
 in the kernel $K(u,x;v,y)$ at $(u,x)=(v,y)$, we adopt 
 the following four-splitting prescription~\cite{Aoki:1990yp}:
 \begin{eqnarray}
   K_{ij}(u,x,v,y) \ \  &\Longrightarrow& \ \   \frac{1}{4} 
       \ [\ K_{ij}(u,x,v_+,y_+) + K_{ij}(u,x,v_+,y_-) \nonumber\\
   & & \ \ \ \ +\  K_{ij}(u,x,v_-,y_+) + K_{ij}(u,x,v_-,y_-)\ ] ,
 \end{eqnarray}
 where    
 \begin{equation}
   v_\pm = \exp\left[V \pm \frac{D_U}{4}\right] , \ \ 
   y_\pm = \exp\left[Y \pm \frac{D_X}{4}\right] . 
 \end{equation}
 Now that all the variables have become discrete and 
 the original integral equation (\ref{eq:IBSeq}) has
 turned into a linear algebraic one, we are able to deal it 
 numerically.

When we solve the IBS equation numerically in this paper, 
we use the following parameters : 
\begin{eqnarray}
&&
 \left[\lambda_U,\Lambda_U \right] = 
  \left[\ 34.5 \times \alpha_\ast - 45.0\
 ,\ \ 
 0.0\ \right], 
\nonumber\\
&&
 \left[\lambda_X,\Lambda_X \right] = 
  \left[\ 43.5 \times \alpha_\ast -52.0\
 ,\ \ 
 0.0\ \right], 
\nonumber\\
&&
 N_{BS,U} = N_{BS,X} = 34.
\end{eqnarray}
Note that 
$\left[ \lambda_U,\Lambda_U \right]$ and 
$\left[ \lambda_X,\Lambda_X \right]$ are chosen so that the dominant
supports of $\Pi_{VV}(Q^2) - \Pi_{AA}(Q^2)$ always lie within the
energy region between UV and IR cutoffs.

%%%%%%%%%%%%%%%%%%%%%%%%%%%%%%%%%%%%%%%%%%%%%%%%%%%%%%%%%%%%%%%%%%%%
%%%%%%%%%%%%%%%%%%%%%%%%%%%%%%%%%%%%%%%%%%%%%%%%%%%%%%%%%%%%%%%%%%%%

\end{document}